\documentclass[twocolumn]{aa}

\usepackage{graphicx}
\usepackage[version=4]{mhchem}
\usepackage{txfonts}
\usepackage{subfigure}
\usepackage{orcidlink}
\usepackage{academicons}
\usepackage{tablefootnote}
\usepackage{gensymb}
\usepackage[normalem]{ulem}
\makeatletter
\renewcommand*\aa@pageof{, page \thepage{} of \pageref*{LastPage}}

\begin{document}

   \title{The MIRI/MRS Library I. Empirically correcting detector charge migration in unresolved sources}

   \author{Danny Gasman
          \inst{1}*\orcidlink{0000-0002-1257-7742}
          \and
          Ioannis Argyriou
          \inst{1}\orcidlink{0000-0003-2820-1077}
          \and
          Jane E. Morrison
          \inst{2}\orcidlink{0000-0002-9288-9235}
          \and
          David R. Law
          \inst{3}\orcidlink{0000-0002-9402-186X}
          \and
          Alistair Glasse
          \inst{4}\orcidlink{0000-0002-2041-2462}
          \and
          Karl D. Gordon
          \inst{3}\orcidlink{0000-0001-5340-6774}
          \and
          Patrick J. Kavanagh
          \inst{5}\orcidlink{0000-0001-6872-2358}
          \and
          Craig Lage
          \inst{6}\orcidlink{0000-0002-9601-345X}
          \and
          Polychronis Patapis
          \inst{7}\orcidlink{0000-0001-8718-3732}
          \and
          G. C. Sloan
          \inst{3,8}\orcidlink{0000-0001-8718-3732}
          }

   \institute{Institute of Astronomy, KU Leuven, Celestijnenlaan 200D, 3001 Leuven, Belgium
         \and
         Steward Observatory, University of Arizona, Tucson, AZ 85721, USA
         \and
        Space Telescope Science Institute, 3700 San Martin Drive, Baltimore, MD, 21218, USA
        \and
        UK Astronomy Technology Centre, Royal Observatory, Blackford Hill Edinburgh, EH9 3HJ, Scotland, UK
        \and
        Department of Experimental Physics, Maynooth University, Maynooth, Co.
Kildare, Ireland
        \and
        Department of Physics, University of California-Davis, 1 Shields Ave. Davis, Ca., U.S.A.
        \and
        Institute of Particle Physics and Astrophysics, ETH Zürich, Wolfgang-Pauli-Str 27, 8049, Zürich, Switzerland
        \and
        Department of Physics and Astronomy, University of North Carolina, Chapel Hill, NC 27599-3255, USA
        \\
            *\email{danny.gasman@kuleuven.be}
             }

   \date{Received 4 April 2024 / Accepted 3 June 2024}

% \abstract{}{}{}{}{} 
% 5 {} token are mandatory
  \abstract
   {The \textit{James Webb} Space Telescope (JWST) has been collecting scientific data for over two years now. The Medium Resolution Spectrometer (MRS) of the Mid-InfraRed Instrument (MIRI) has been one of the telescope's most popular modes, and has already produced ground-breaking results. Scientists are now looking deeper into the data for new exciting discoveries, which introduces the need to characterise and correct known systematic effects to reach the photon noise limit. Five important limiting factors for the MRS are the pointing accuracy, non-linearity, detector charge migration, detector scattering---resulting in both spatial broadening and spectral interferometric fringing---the accuracy of the point-spread function (PSF) model, and the complex interplay between these.}
   {The Cycle 2 calibration programme 3779, entitled ‘The MIRI/MRS Library', proposed a 72-point intra-pixel dither raster of the calibration star 10-Lac, which provides a unique dataset tailored for the purpose of addressing the limiting factors on the MRS data accuracy. In this first work of the paper series, we aim to address the degeneracy between the non-linearity and charge migration (brighter-fatter effect) that affect the pixel voltage integration ramps of the MRS. Due to the low flux in the longer wavelengths, we only do this in the 4.9 to 11.7~micron region (spectral channels 1 and 2).}
   {We fitted the ramps individually per pixel and dither, in order to fold in the deviations from classical non-linearity that are caused by charge migration. The ramp shapes should be repeatable depending on the part of the PSF that is sampled. By doing so, we defined both a grid-based linearity correction, and an interpolated linearity correction.}
   {Including the change in ramp shape due to charge migration yields significant improvements compared to the uniform illumination assumption that is currently used by the standard JWST calibration pipeline. The standard deviation on the pixel ramp residual non-linearity is between 70-90\% smaller than the current standard pipeline when self-calibrating with the grid. We are able to interpolate these coefficients to apply to any unresolved source not on the grid points, resulting in an up to 70\% smaller standard deviation on the residual deviation from linearity. After applying the correction, the full-width at half maximum is up to 20\% narrower for sources that cover the full pixel dynamic range. Furthermore, the depth of the fringes is now consistent up the ramp, improving the standard deviation on the difference in fringe depth between the start and ends of integrations by $\sim$60\%.}
   {Pointing-specific linearity corrections allow us to accurately model the pixel ramps across the PSF, and for the first time, fix the systematic deviation in the slopes. In this work we demonstrated this for unresolved sources. The discovered trends with PSF sampling suggest that, in the future, we may be able to model ramps for spatially extended and resolved illumination as well.}

   \keywords{ Astronomical instrumentation, methods and techniques – Instrumentation: detectors – Methods: data analysis – Methods: numerical – Infrared: general }

    \titlerunning{The MIRI/MRS Library I}
    \authorrunning{D. Gasman et al.}
   \maketitle

%
%-------------------------------------------------------------------

\section{Introduction}
\label{sec:introduction}

The Medium Resolution Spectrometer \citep[MRS;][]{ref:15WePeGl,ref:23ArGlLa} of the Mid-InfraRed Instrument \citep[MIRI;][]{ref:23WrRiGl} on board the \textit{James Webb} Space Telescope \citep[JWST;][]{ref:23RiPeMc,ref:23GaMaAb} has allowed us to once more peer into the mid-infrared, with improved sensitivity and resolution. After the first two years of operation, the instrument has produced many stunning results in a variety of fields, from galaxies, to young dusty stars with disks, to brown dwarfs and exoplanets.

Still, there are a multitude of exciting science cases that are currently affected and limited by instrument effects. The characterisation and identification of faint molecular features is affected by the presence of residual fringing (Kavanagh et al. in prep.), especially for bright unresolved and semi-extended sources, and the potential altering of specific features by the standard JWST pipeline fringe corrections \citep{ref:23GaArSl}. Furthermore, disentangling faint companions from the central star via point-spread function (PSF) subtraction is complicated by the current limited understanding of the PSF and charge migration (Patapis et al. in prep.). Therefore, there are a multitude of factors that are currently preventing us from pushing the instrument to its limits. These factors include (1) the repeatability of the pointing, (2) the non-linearity of the pixel ramps, (3) detector charge migration, (4) scattering within the detectors (resulting in spatial broadening and spectral fringing), and (5) the accuracy of the PSF model. The uncertainties surrounding these aspects impact the accuracy in absolute flux, the accuracy in optimised spectral extraction methods such as PSF-weighted spectrophotometry, and the contrast that may be reached when subtracting the PSF.

\begin{figure}[h!]
    \centering
    \includegraphics[width=\columnwidth]{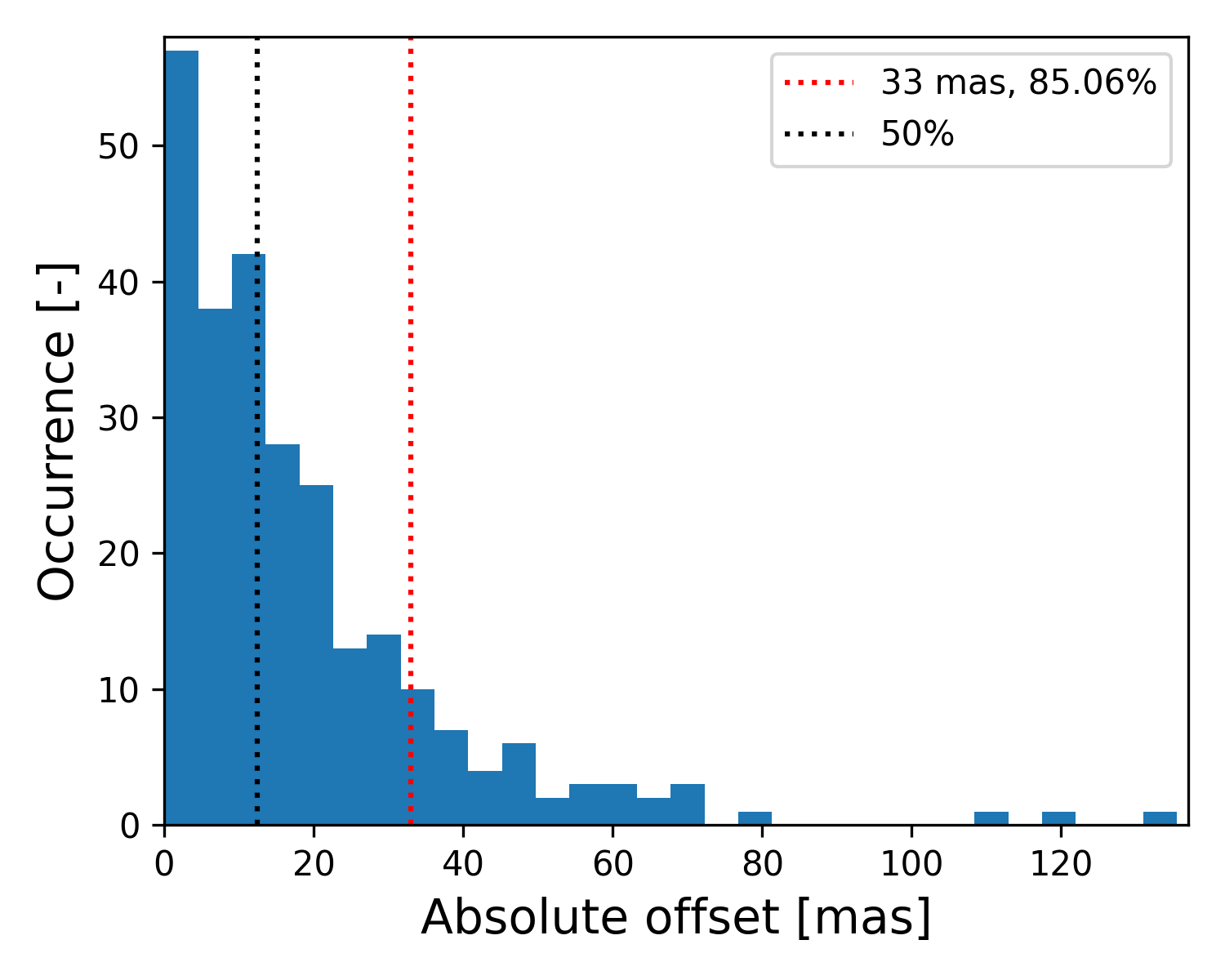}
    \caption{Histogram of MRS pointing errors based on offsets between measured and nominal centroids for point sources observed with target acquisition in commissioning and Cycle 1.}
    \label{fig:pointing}
\end{figure}

In Fig.~\ref{fig:pointing} we show the astrometric precision of the MRS based on statistics from commissioning and Cycle~1 observations of point sources that used Target Acquisition (TA). It illustrates the ability of the MRS to place the source in the targeted locations (see also the discussion by \citealt{ref:24PaArLa}). For reference, the size of an MRS pixel is approximately 200~mas. \citet{ref:23GaArSl} show that these small sub-pixel offsets cause a large enough change in the fringe properties (amplitude and phase) to result in 10\% residuals when using a reference star spectrum. This then affects the absolute flux calibration in the spectra and extracted science. The red dotted vertical line at 33~mas shows that 85\% of observations fall within that limit. A three-by-three intra-pixel raster with sides of length 66~mas would therefore allow us to characterise and correct for some of the effects linked to the pointing non-repeatability.

\begin{figure}[]
    \centering
    \includegraphics[width=\columnwidth]{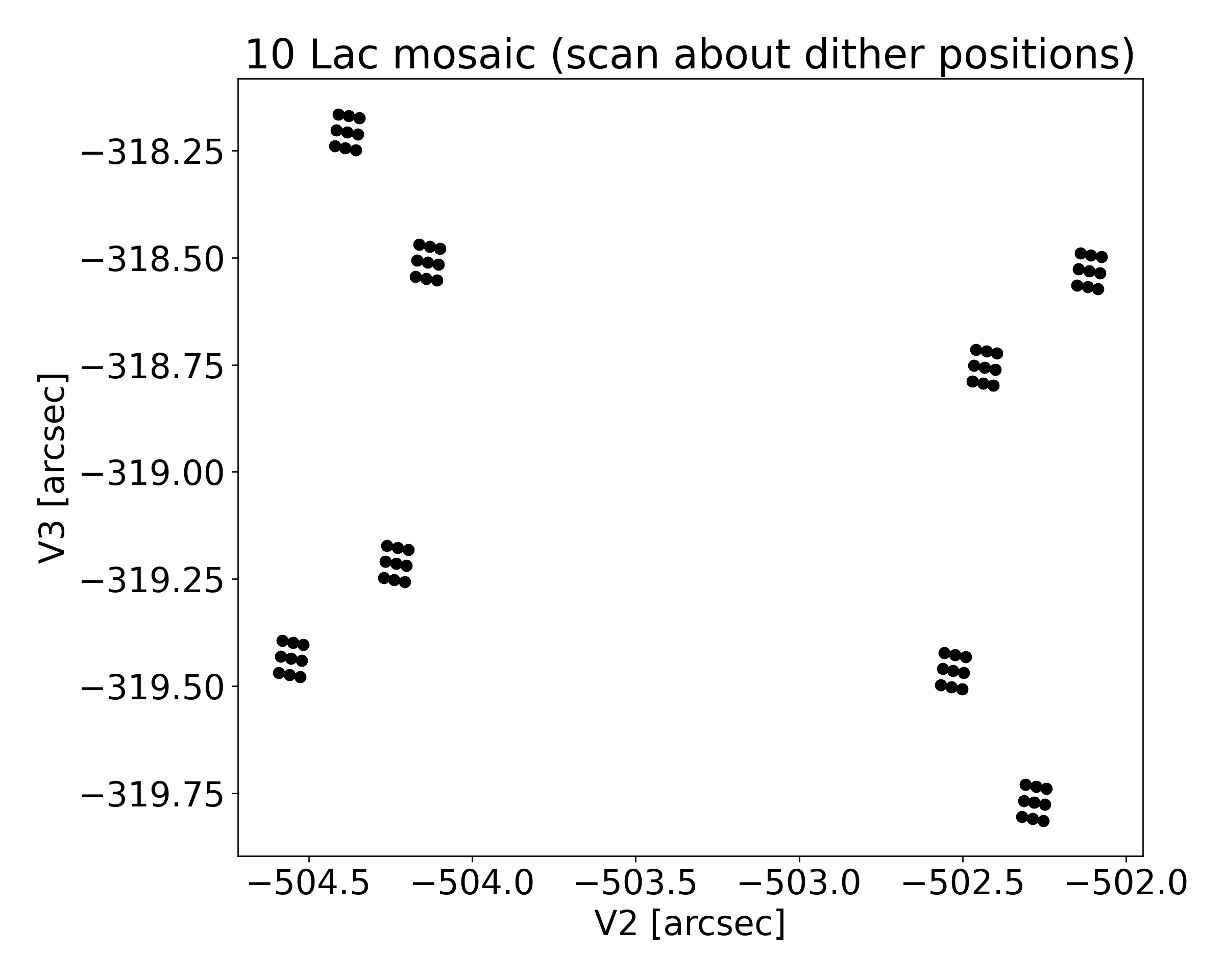}
    \caption{Set-up of the mosaic in programme 3779 around the \texttt{4-POINT POSITIVE} and \texttt{4-POINT NEGATIVE} dither positions.}
    \label{fig:mosaic}
\end{figure}

The JWST Cycle 2 calibration programme PID~3779 (‘The MIRI/MRS Library'; PI: D. Gasman, co-PI: I. Argyriou; \citealt{ref:23GaArGl}) aims to assess and correct the systematic uncertainties introduced by the aforementioned issues on the data. In order to do so, we targeted the O9 star 10-Lac\footnote{From SIMBAD} ($K$ magnitude 5.5) and performed a three-by-three intra-pixel mosaic around all nominal point source dither positions; \texttt{4-POINT POSITIVE} and \texttt{4-POINT NEGATIVE} in the Astronomer's Proposal Tool (APT). This target was selected since its properties are well known from the previous astrometric calibration scan described in \citet{ref:24PaArLa}. The commanded on-sky positions are shown in Fig. \ref{fig:mosaic}, and the comparison between the planned and real positions can be found in Fig.~\ref{fig:observed dithers}.\footnote{The work to investigate the small difference in commanded versus acquired pointing falls outside the scope of this work.} While not entirely in the same location on-sky, the programme demonstrates the impressive pointing accuracy and stability of JWST, allowing for this unconventional set-up. The length of the exposure integrations is 100 groups, in order to reach saturation and infer the full ramp shape up to this point in channel 1. Using \texttt{FASTR1} read-out, the pixel integration ramps are finely sampled. Reaching saturation becomes more difficult for longer wavelengths, however, due to the reduced stellar flux (see the 10-Lac spectrum in Fig.~3 in \citet{ref:23LaMoAr}), and the lower quantum efficiency of the MRS long wavelength detector above 20~$\mu$m. For this reason, the dynamic range reached in channels 3 and 4 is not sufficient to generate linearity corrections. The dynamic range probed by the ramps per band is shown in Fig.~\ref{fig:range_per_band}.

The different goals of the programme are divided into four different papers, following a bottom-up approach. Here, Paper I, we start on the detector ramp level (stage~1 of the JWST calibration pipeline \citep{ref:23BuEiDe}). We aim to characterise the degeneracy between non-linearity and charge migration \citep[brighter-fatter effect (BFE),][]{ref:23ArLaRi} in the presence of large signal contrasts caused by the PSF and the spectral fringes (more details regarding the specifics of these effects can be found later in this introduction), in order to be able to correct for this effect. In Paper II, the focus is on the slopes derived from the corrected ramps (after stage 1). These slopes give the rate of photo-electron accumulation, which is proportional to the incoming astrophysical flux. We use the slopes to re-assess the fringe pattern and spectrophotometric calibration when dealing with intra-pixel shifts in the PSF center. The goal of Paper III is to create detector-based PSF models and examine the potential of detector-based spectral extraction and PSF subtraction on the detector image plane, to benefit from the full spatial and spectral resolution of the instrument. Finally, in Paper IV we will turn to resolved sources, rather than unresolved sources, to assess the potential of forward modelling the fringe pattern (see the rest of this section for details regarding fringing, and e.g. Sect~4.5 of \citealt{phdthesisYannis} and \citealt{ref:22GaArVa} for examples of the forward modelling approach).

In this work, we focus on the detector ramps. The MRS uses two arsenic-doped silicon impurity band conduction (Si:As IBC) detectors \citep{ref:15RiReMo} which record four spectral channels (1-4) with each MRS exposure \citep{ref:15WePeGl}. Each of these channels contains three sub-bands, which we refer to as A (SHORT), B (MEDIUM), and C (LONG) in this work. The local spatial detector coordinates are $\beta$, which are constant per slice of the image slicer of the Integral Field Units, and $\alpha$, which vary within a slice \citep{ref:24PaArLa}. Detector pixels record signal by sampling ‘up-the-ramp', which yields Data Numbers ($DN$) versus groups (in the case of MIRI there is 1 frame per group, and for FASTR1 one frame takes 2.755~s to read out, \citealt{ref:15ReSuFr}). The photoelectrons are generated in the detector infrared-active layer, and subsequently attracted to the frontside contact of the associated pixel due to the field created by the voltage difference (the bias voltage) between the frontside and buried contacts \citep[see][Fig. 1]{ref:23ArLaRi}. 

\begin{figure}[t]
    \centering
    \includegraphics[width=1\columnwidth]{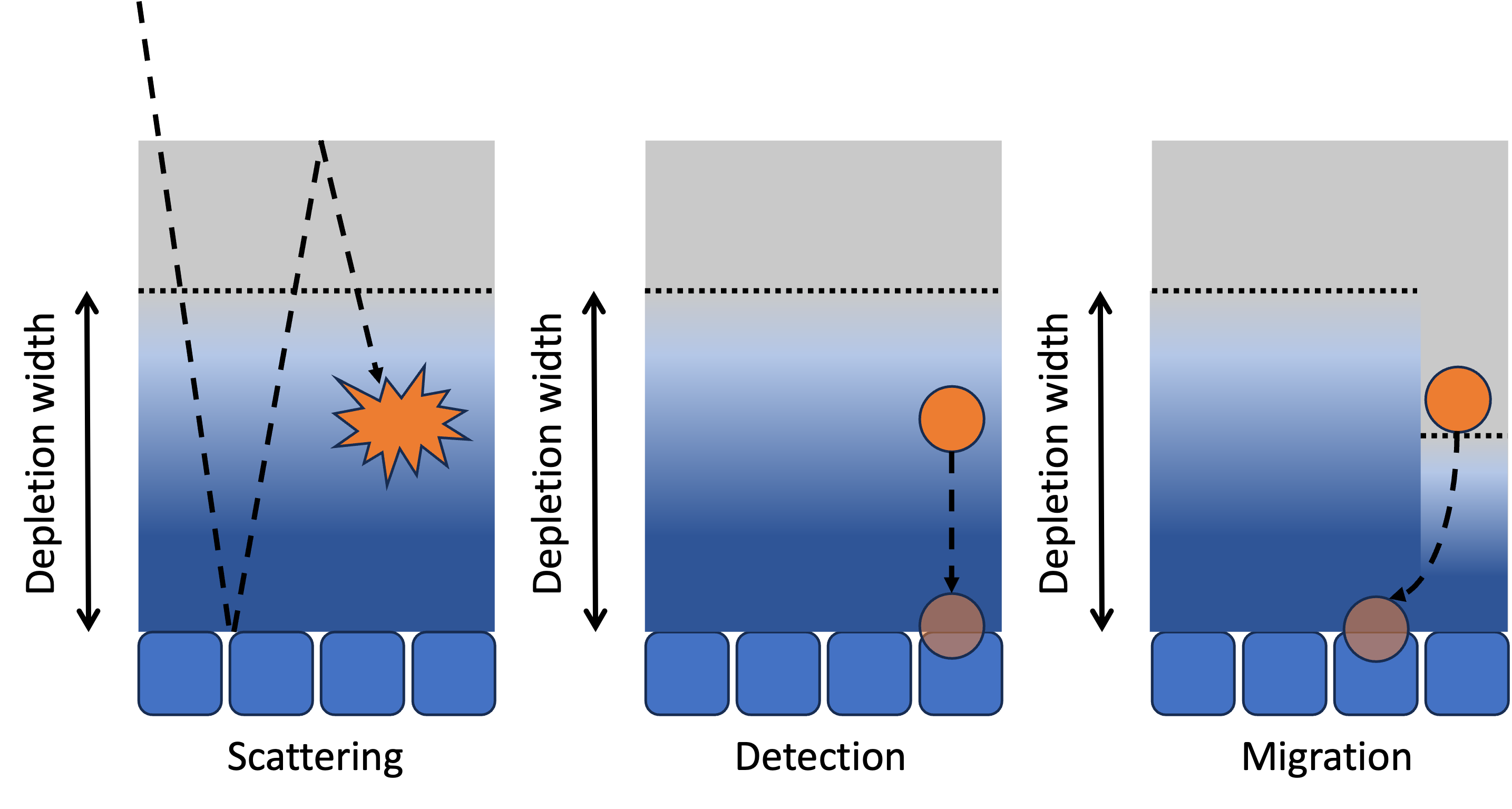}
    \caption{Scattering (spatial fringing, left panel), direct detection (ideal case, central panel) and migration (BFE, right panel) in the detector. The blue gradient signifies the depletion width. The squares are the pixels, and the circles depict photo-electrons.}
    \label{fig:migration_schematic}
\end{figure}

As charge is accumulated by a pixel, the bias voltage across the infrared-active layer above that pixel decreases, causing the effective depletion width to shrink. Therefore, the effective response of the detectors decreases as charge is accumulated, resulting in non-linear ramps \citep{ref:23MoDiAr}. We consider this to be the ‘classical non-linearity', applicable when illuminating all detector pixels uniformly. However, the reducing depletion width has a second effect on the ramps. In cases of large contrast in bias voltage between neighbouring pixels caused by contrast in detector illumination (e.g. for a point source), the depletion widths of two neighbouring pixels can greatly differ. Due to this, a less illuminated pixel with a correspondingly larger remaining depletion width can attract photoelectrons generated in a brighter neighbouring pixel with a much smaller remaining depletion width. In Fig.~\ref{fig:migration_schematic} the charge migration is illustrated in the right-most panel. This BFE is addressed in more detail in \citet{ref:23ArLaRi} and Gasman et al. (in prep), and is similar to the BFE typically observed in near-infrared detectors \citep[e.g.][]{ref:17LaBrTy,ref:18CoArSm,ref:18PlShSm,ref:20HiCh}. As modelled in \citet{ref:23ArLaRi}, the ramp shapes can change dramatically in cases of contrast between neighbouring pixels, resulting in incorrect estimates of the flux. Especially at the short MRS wavelengths, the undersampling of the PSF results in a large signal contrast between pixels. Due to the charge spilling to neighbouring pixels, the BFE can broaden the PSF up to 20\% in the MIRI Imager \citep{ref:23ArLaRi,ref:23MoDiAr}.

\begin{figure}[t]
    \centering
    \includegraphics[width=1\columnwidth]{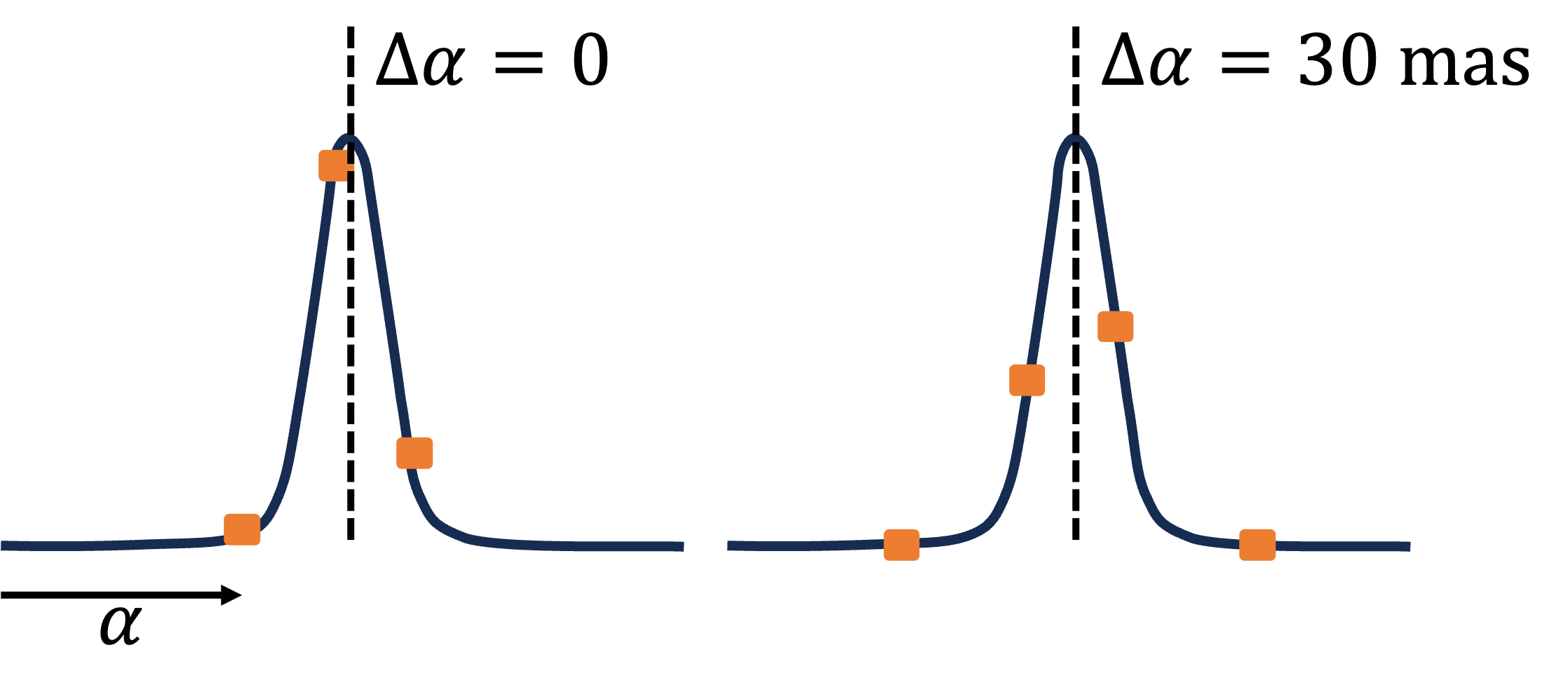}
    \caption{Change in sampling of the PSF with $\alpha$ offset. The orange markers indicate the parts sampled by the pixels.}
    \label{fig:sampling}
\end{figure}

\begin{figure*}
    \centering
    \includegraphics[width=1\textwidth]{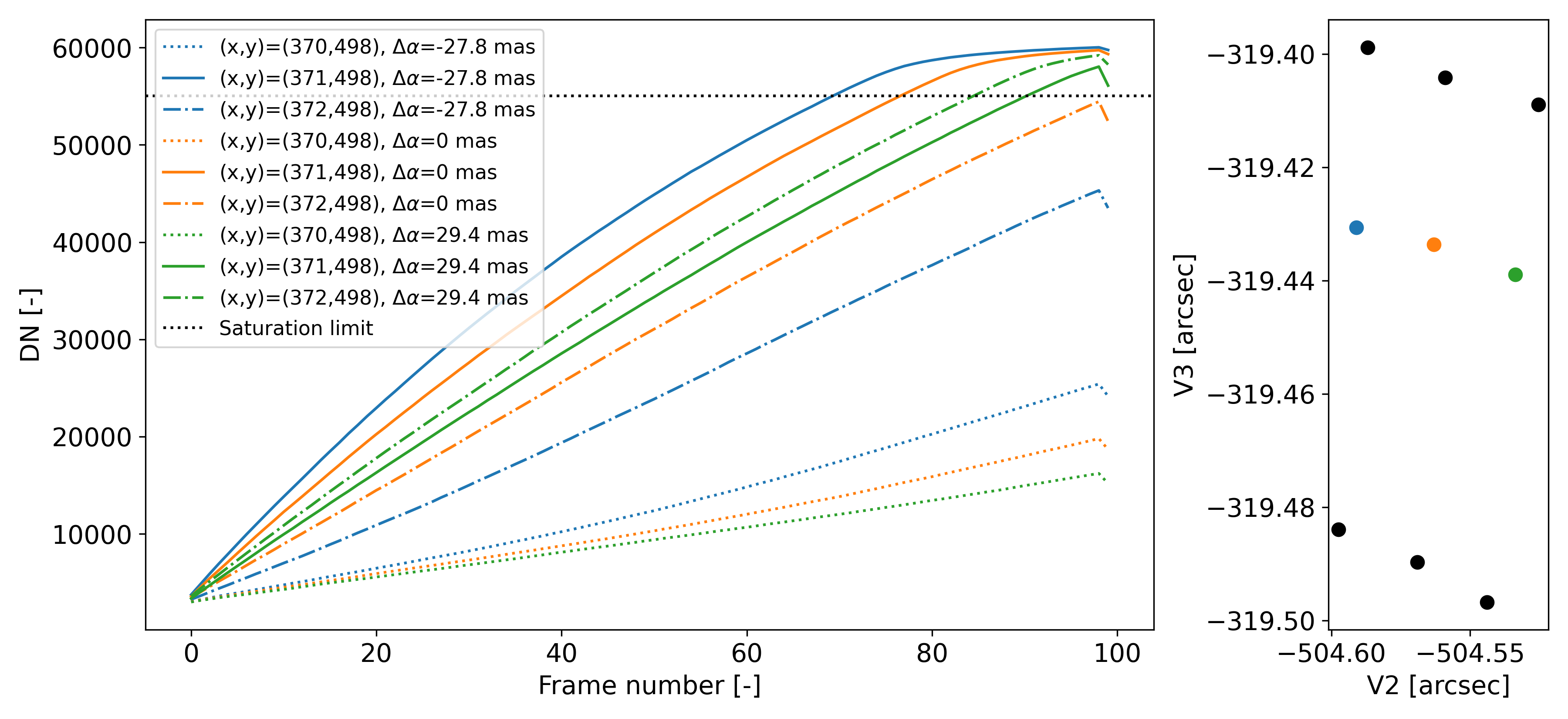}
    \caption{Changing ramp shapes with sub-pixel offsets in band 1A (left). Each linestyle corresponds to a different pixel on the detector (ranging from PSF peak to wings), with the $x,y$ pixel coordinates given in the legend. Each colour corresponds to a different pointing, with the offsets in $\alpha$ with respect to the central pointing given in the legend. Only three samples of the nine-point raster are shown for clarity. The panel on the right shows which these are in the JWST V2/V3 coordinates.}
    \label{fig:bfe_contrast}
\end{figure*}

The contrast-dependent systematic on the ramp shape is not taken into account in the non-linearity correction in the JWST calibration pipeline for the MRS, which is a limitation of the current pipeline. We can now forgo this assumption and improve upon the calibration methods. Since high debiasing contrast between pixels can affect the ramp shapes, in order to determine the ‘true' non-linearity of the MIRI ramps, observations of completely uniform illumination would be required to derive the correction. However, there is no such thing as ‘flat' illumination for the MRS. Due to the slit masks of the MRS integral field unit, only traces are illuminated on the detector, with a varying transmission within the profile of each trace (i.e. the illumination is not uniform). Furthermore, due to the reflectivity of the different layers within the detectors, photons not detected at first pass due to the relatively low absorption efficiency of arsenic in channels 1 and 2 \citep{ref:21GaRiGu} scatter through the detector between the anti-reflection coating and the pixel metallisation. Fig.~\ref{fig:migration_schematic} illustrates this in the first panel, where a photon is free to scatter through the detector prior to generating a photo-electron and being detected. While this results in broadening of the PSF in the spatial direction \citep{ref:23ArGlLa}, the photons produce an interference pattern in the spectral direction. This constructive and destructive interference results in a sinusoidal pattern on the detector. This is referred to as fringing; more details about this effect on point source observations can be found in \citet{ref:20ArWeGl} and \citet{ref:23GaArSl}. Detector interferometric fringing naturally results in alternating brighter and fainter illumination with a peak-to-trough distance as low as four pixels, creating a contrast of up to 20\% at 5~$\mu$m. For this reason, the non-linearity correction in the current pipeline is based on fringe peaks in an extended source for the MRS, as these ramps most closely resemble the classical non-linearity shape \citep{ref:23MoDiAr}. For all other cases this solution is not representative. In the illumination case we are interested in, that of the point source, depending on how the source lands on the detector and the sampling of the PSF changes, this contrast is increased or decreased, therefore changing drastically between small sub-pixel offsets. This is illustrated in Fig.~\ref{fig:sampling}, and the changes in ramp shapes with offset in the grid can be found in Fig.~\ref{fig:bfe_contrast}. Depending on the contrast, the pixels that sample the wings of the PSF show the characteristic convex BFE shape (e.g. the dotted curves), as opposed to the concave PSF centre (e.g. solid curve). Clearly, the ramp shapes change with the sub-pixel changes in PSF sampling. When the source is dispersed on the same part of the detector, the ramp shapes will be the same, as is the case for fringing \citep{ref:23GaArSl}. Therefore, we can use the grid of reference point sources from PID~3779 to derive pointing specific non-linearity solutions that fold in the changes in contrast and find a more fitting solution for all MRS point source observations.

The work presented here is structured as follows. In Sect.~\ref{sec:methods} the methods used to linearise the ramps are presented. These new corrections are tested in Sect.~\ref{sec:results}, where we give the statistics after linearisation, assess the possibility to interpolate between grid positions, and present the improvements on the full-width at half maximum (FWHM) for unresolved sources that cover the full dynamic range (although non-saturating observations also benefit from this correction). The caveats are discussed in Sect.~\ref{sec:discussion}, and our findings are summarised in Sect.~\ref{sec:conclusion}.

% Persistence?

\section{Methods}
\label{sec:methods}

\subsection{Data reduction}
\label{sec:data_reduction}
This work made use of the JWST calibration pipeline version 1.12.5 \citep{ref:23BuEiDe}, assuming CRDS \texttt{pmap} 1197. Prior to fitting the ramps to derive the correction curves, first the ramps were calibrated based on the following considerations. (1) Aside from the linearity correction, one other step currently impacting the ramp shape is the dark current correction step: \texttt{dark\_current} \citep{ref:23MoDiAr}. Therefore, in order to ensure consistency with the JWST pipeline, and make sure the linearity correction linearises the ramps even when applying this step of the JWST pipeline, the fits were performed after running this step on the uncalibrated data. (2) A number of frames at the start of the ramps are affected by the reset effects \citep{ref:23MoDiAr}. While a part of this is mitigated by the dark correction, which now incorporates the \texttt{reset} step, it is not yet well-characterised, hence we skipped the first four frames to limit its effects. (3) The saturation threshold in the MRS detectors is set at 55,030~$DN$.\footnote{See the appropriate reference files at \url{https://jwst-crds.stsci.edu/}.} To prevent saturation from being included in the fit, samples above that $DN$-level were omitted from the fit. (4) Cosmic ray hits cause jumps in the ramps. These jumps will affect the fit and reduce its accuracy, and even after a cosmic ray hit, it can take a number of frames for the ramp to settle down again \citep[see][Fig. 17]{ref:23MoDiAr}. Therefore, ramps with jumps where the resulting error on the fit was larger than the noise floor level of about 50~$DN$ were ignored, and the neighbouring linearity correction values were spliced in, unless the jump happens very late in the integration ($>$80 frames), in which case the ramp was fitted up to the jump. We refer the reader to Fig.~\ref{fig:sinusoid} in Sect.~\ref{sec:discussion} for a visualisation of the noise level that motivated our choice.

\subsection{Ramp linearisation in the presence of BFE}
\label{sec:bayes}
\begin{figure}[h!]
    \centering
    \includegraphics[width=0.8\columnwidth]{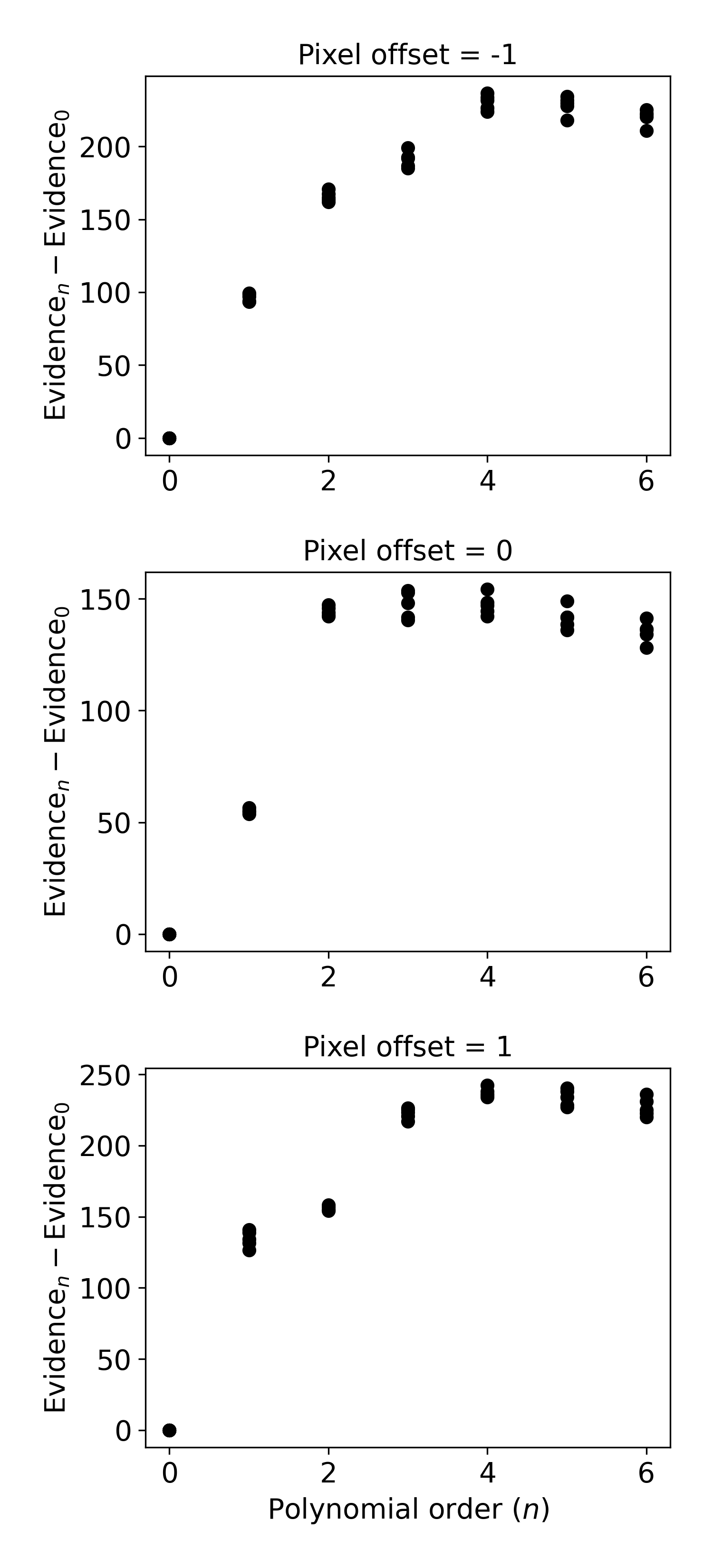}
    \caption{Bayesian evidence of increasing polynomial order fits for groups of ramps with similar shapes. Here ramps on fringe peaks are shown, either in the brightest PSF sample or with one pixel offset to the left (negative, top panel) or right (positive, bottom panel) on the detector. The evidence parameter was normalised by subtracting the evidence of the zeroth order fit.}
    \label{fig:bayes_peaks}
\end{figure}

In order to derive the linearity correction curves, first a polynomial was fitted to the ramps. This fit is a polynomial of order $n$ with coefficients $c$ as a function of time $t$:
\begin{equation}
    DN_{fit} (t) = \sum_{i=0}^n c_i \cdot t^i\text{.}
\end{equation}
The non-linearity correction in the current pipeline is based on the average of the ramps in fringe peaks in an extended source for the MRS \citep{ref:23MoDiAr}. As demonstrated in Fig.~\ref{fig:bfe_contrast}, the ramp shapes in the wings of the PSF are different from the shapes in the PSF centre. Therefore, when applying this constant correction to the less conventional shapes of BFE-affected ramps the corrected ramp will not be linear. First, in order to examine the most optimal polynomial order to use for the different pixel ramp shapes in point source observations, we used a Bayesian evidence routine to prevent over-fitting. Using the PYTHON package \texttt{BayesicFitting} \citep{ref:21KeMu}, we fitted polynomials of increasing order to the ramps, until less evidence was found to choose the higher order model over a lower order model. This would mean the evidence plateaus or even decreases after a certain polynomial order. For this analysis we combined ramps that are (1) in the fringe peaks and PSF peak defined as ‘pixel offset = 0', and (2) in the fringe peaks and in the PSF wings, defined as ‘pixel offset = $\pm$1'. The results of this analysis are presented in Fig. \ref{fig:bayes_peaks}. For the PSF wings the evidence plateaued around polynomial orders of 3, or even 4. On the other hand, in the PSF peak the polynomial orders 2 or 3 showed the highest evidence.

Since the ramp shapes change quickly between pixels, each pixel will have its own associated linearity correction. Indeed, when applying corrections of the optimal orders the residuals are the smallest. However, the noise in the resulting spectrum was much larger, which we demonstrate in Fig.~\ref{fig:noise}. When using the maximum polynomial order of 4, the resulting rates showed much more spurious scatter around the sinusoidal fringe pattern when assuming a third order polynomial for all ramps. In the wings of the PSF the S/N in the ramps reduced since fewer photons are detected, which likely caused noisier fits when using a higher order polynomial. As a compromise between linearisation and limiting the noise in the resulting spectrum, we fitted a third-order polynomial for all ramps. The resulting statistics of these fits on linearising the ramps (corresponding to flux accuracy) can be found in Sect.~\ref{sec:stats}.

After fitting the third-order polynomials to the complete ramps (processed as per Sect.~\ref{sec:data_reduction}), correction curves were calculated. The correction curves were derived using the same method as in \citet{ref:23MoDiAr}, although in this work this is done separately per pixel, instead of an average of all ramps. If the illumination is repeatable, the ramp shapes should be constant and independent of flux, and the ramp is corrected using the right assumed shape. First, the linear part of the third-order polynomial fit was assumed to be the true ramp shape, in other words the true accumulated photoelectric charge rate from the astrophysical source for an ideal detector suffering no electronic systematic effects:
\begin{equation}
    DN_{linear} (t) = c_0 + c_1 \cdot t\text{.}
\end{equation}
In order to derive how much the ramps deviated from the true ramp shape, the observed $DN_{obs}$ was divided by the assumed linear solution $DN_{linear}$. This was then plotted against the $DN_{obs}$ instead of time, since the deviation from linearity depends on this measure. This is due to the relation to the measure of debiasing from the nominal bias at a certain pixel \citep{ref:23ArLaRi}. The non-linearity correction for the specific pixel can be derived by fitting the resulting curve:
\begin{equation}
    (DN_{linear}/DN_{obs}) (t) = \sum_{i=0}^{n} L_i \cdot DN_{obs}^i (t)\text{,}
    \label{eq:corr_curve}
\end{equation}
where $L_i$ are the fit coefficients and are further referred to as correction coefficients. At the start of the ramp (time step 0~s), which is not recorded by the read-out electronics, the $DN$-level is at an arbitrary video offset (a bias $DN$-level). In addition to this, the linear component of the fit should be equal to 1, as the flux of the object (linear term) should not change. Therefore, $L_0$ was set to 0, and all coefficients were normalised to $L_1$, to ensure that $L_1=1$. The pipeline then substitutes the $DN$-values per frame in the correction curve, described by the result of fitting Eq.~\ref{eq:corr_curve}, resulting in the final linearised ramp.

The corrections were derived for pixels up to $\pm$2 detector columns (X-pixels) away from the brightest pixel, and $\pm$1 integral field slice number away from the brightest slice \citep{ref:24PaArLa}. The ramps farther from the PSF centre were too faint, and could not be used. Therefore, for all other pixels, we assumed the current pipeline correction, which should be accurate enough for these faint pixels where the ramps are still largely linear, as the $DN$-level reached is low.

\section{Results}
\label{sec:results}

\begin{figure*}
	\centering
    \includegraphics[width=0.9\textwidth]{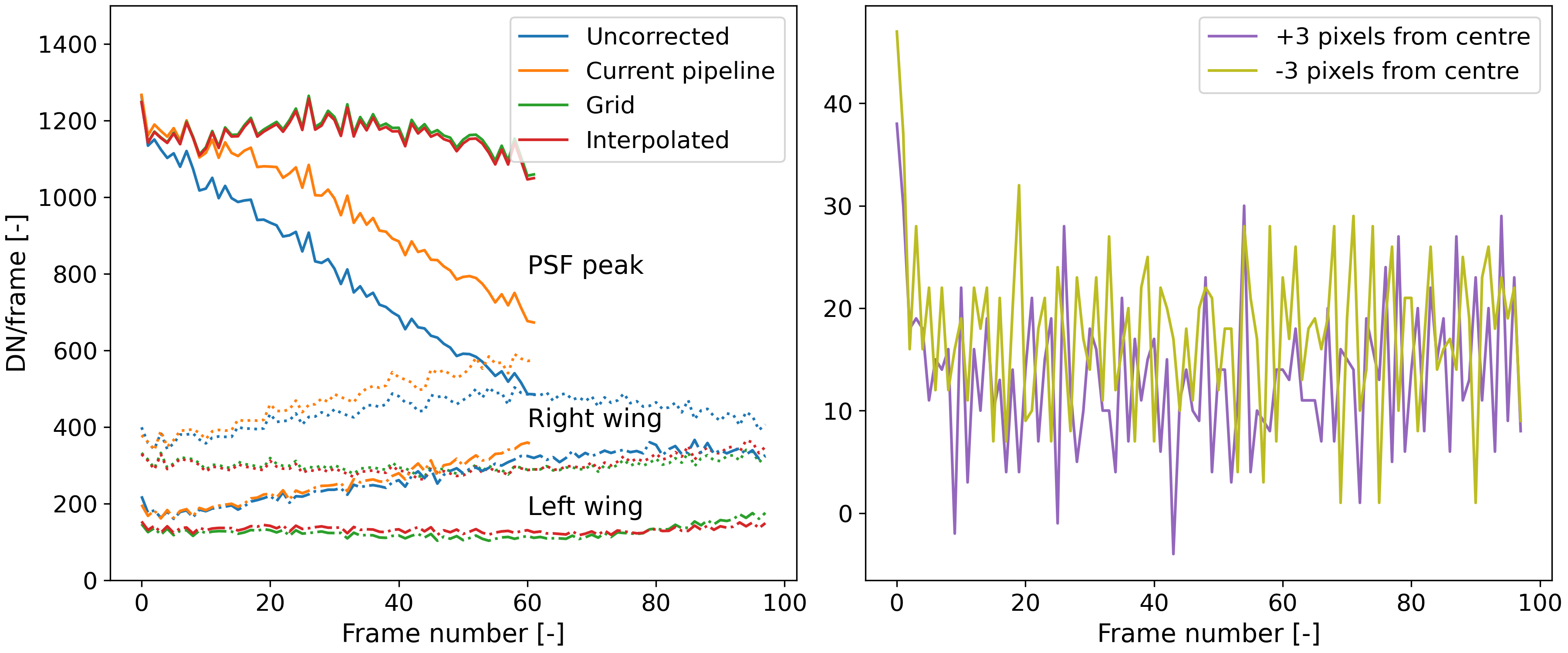}
    \caption{Frame differences of the ramps for the different calibration methods in the PSF centre ($(x,y)=(415,940)$), and the PSF wings ($(x,y)=(414,940)$ and $(x,y)=(416,940)$), in band 1A (left). To better visualise the deviation from perfect linearity, we show the subsequent frame difference ($DN_{i+1}-DN_{i}$). A perfectly linear ramp should be horizontal and straight. Note that the ramps in the panel on the left are cut at the saturation point. We also show the frame differences of the uncorrected ramps 3 pixels from the PSF centre (right).}
    \label{fig:linearised_ramps}
\end{figure*}

\begin{figure*}
	\centering
    \includegraphics[width=0.7\textwidth]{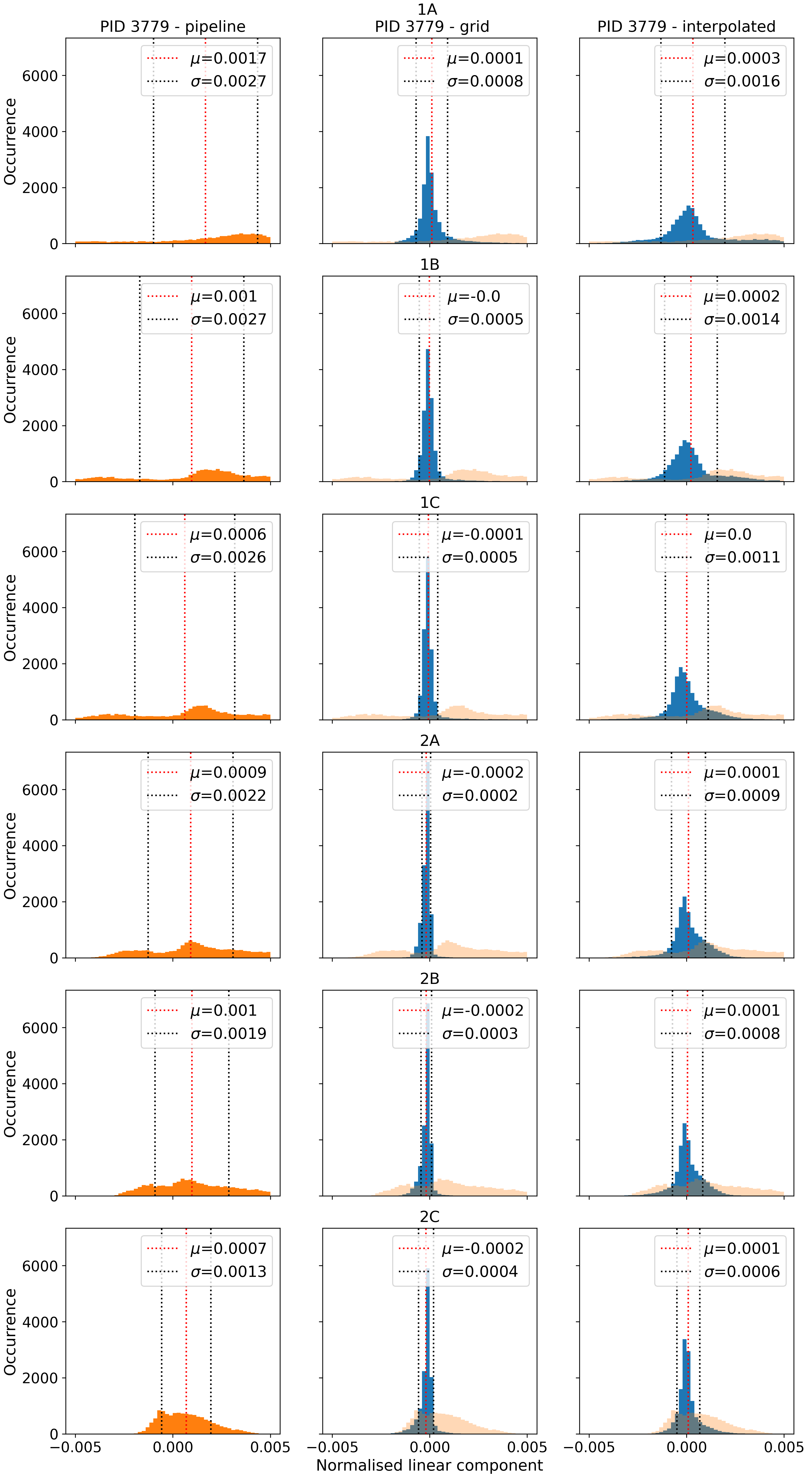}
    \caption{Remaining linear component in a subset of the PID~3779 data after correcting the ramps using the current pipeline files (left), the grid directly (middle), or the interpolated coefficients (right).}
    \label{fig:stats_self}
\end{figure*}

\subsection{Linearity statistics}
\label{sec:stats}
First, we demonstrate the qualitative improvements in ramp linearisation in Fig.~\ref{fig:linearised_ramps}, where we show a comparison between the uncorrected ramps, the current pipeline, and the solutions presented in this work in the left panel. The right panel demonstrates that ramps three pixels from the PSF centre do not deviate significantly from linearity even when not applying a correction, due to their low signal, at least in channel 1. We note that the linearisation by the current pipeline is not ideal, even in the PSF peak. Even more significant are the PSF wings, where the correction is in fact in the wrong direction, increasing the rate of the ramps even further beyond the BFE contribution. This is accounted for with the corrections presented in this work. The interpolated solution will be discussed in more detail in Sect.~\ref{sec:interp}, but below we examine the statistics of directly applying the grid.

We run the standard JWST calibration pipeline, in other words, the averaged correction derived from the fringe peaks with a fully extended illumination, and the new pointing-specific point source corrections. Then, for each ramp that traces the PSF, we find the difference per subsequent frame and divide by the mean of the frame differences $\left (\frac{DN_{i+1}-DN_{i}}{\overline{\Delta DN}} \right)$. A linear regression is fitted through this. The linear component should be 0 if the ramp has been perfectly linearised, with a Gaussian spread ideally only dependent on read, thermal, and photon noise. 

We first check the results on the data themselves. A subset of four dithers is used to plot the statistics. The normalised linear component of all ramps following the PSF trace (central pixel up to $\pm$2 pixels) has been plotted per band in Fig.~\ref{fig:stats_self}. We first focus on the left and central columns. When not including the influence of contrast, the spread in the remaining linear component is large, indicating that the ramps are not consistently linear, which has a direct impact on the resulting slope and derived flux uncertainty of the pixel. Moving to central column, the linear components are centred around 0, with a much smaller standard deviation: up to 70-90\% smaller. As we increase in wavelength (moving from the top to bottom rows in the figure, from band 1A to 2C) the current pipeline solution starts to converge towards the centre as well. This is due to the lack of dynamic range reached; as shown in Fig~\ref{fig:range_per_band}, the pixels are not reaching high $DN$ values, hence the non-linearity is not as significant (at low $DN$ the detector response is close to linear).

\begin{figure*}
	\centering
    \includegraphics[width=0.7\textwidth]{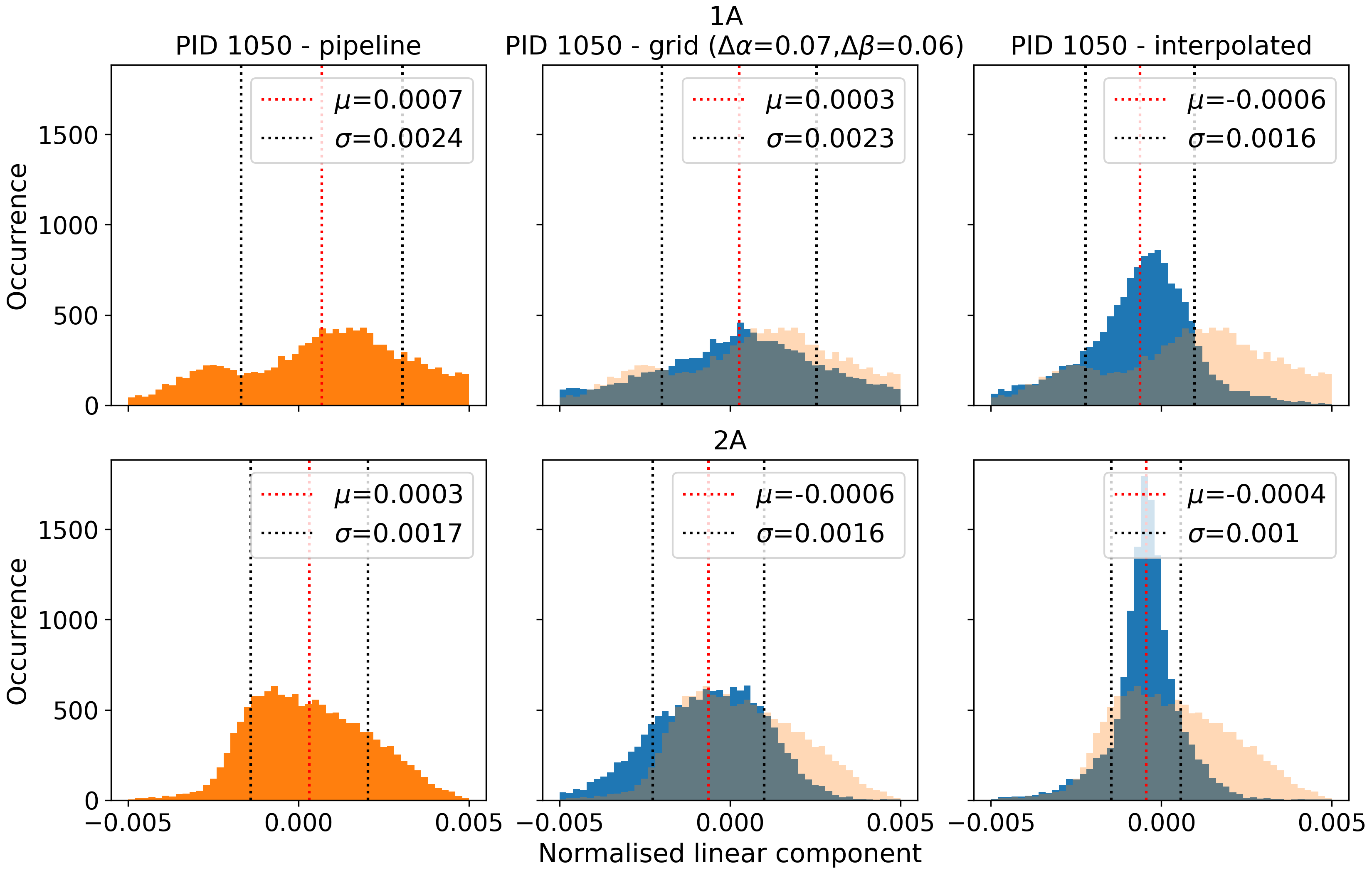}
    \caption{Remaining linear component in observation 7 of PID~1050 after correcting the ramps using the current pipeline files (left), the grid directly (middle), or the interpolated coefficients (right). The top row shows band 1A, and the bottom row band 2A.}
    \label{fig:stats_ref}
\end{figure*}

Next, the new correction is tested on a different target. In order to be able to perform this test, the test target needs to reach at least $\sim$30,000-40,000~DN, and be observed with target acquisition in one of the point source dither patterns. This yielded one publicly available observation: HD~158485 from PID~1050 (PI: B. Vandenbussche). As this was a cross-dichroic test, only the observations of sub-band A have the correct set-up. The results of this test can be found in Fig.~\ref{fig:stats_ref}. Applying the grid directly to this observation does not yield a significant improvement compared to the current pipeline. This is due to the centroid of the observations falling outside of the three-by-three grid shown in Fig.~\ref{fig:pointing}, and the offsets are around 60-70~mas in both spatial dimensions for all dithers. The average of all the offsets of the four different dithers taken together are noted at the top of the central column. Therefore, if an observation falls outside the grid points, directly using these corrections does not improve the linearity of the ramps. However, by combining 72 dithers, we have a well-sampled PSF per row. In Sect.~\ref{sec:interp} we assess the possibility of interpolating the grid point corrections to find a solution based on how the PSF is sampled in $\alpha$-space, in order to generate an approximate reference for observations that are too far outside the grid. However, we first assess the influence of charge migration on spectral fringing.

\subsection{Charge migration and fringing}
\label{sec:contrast}

\begin{figure*}
    \centering
    \includegraphics[width=\textwidth]{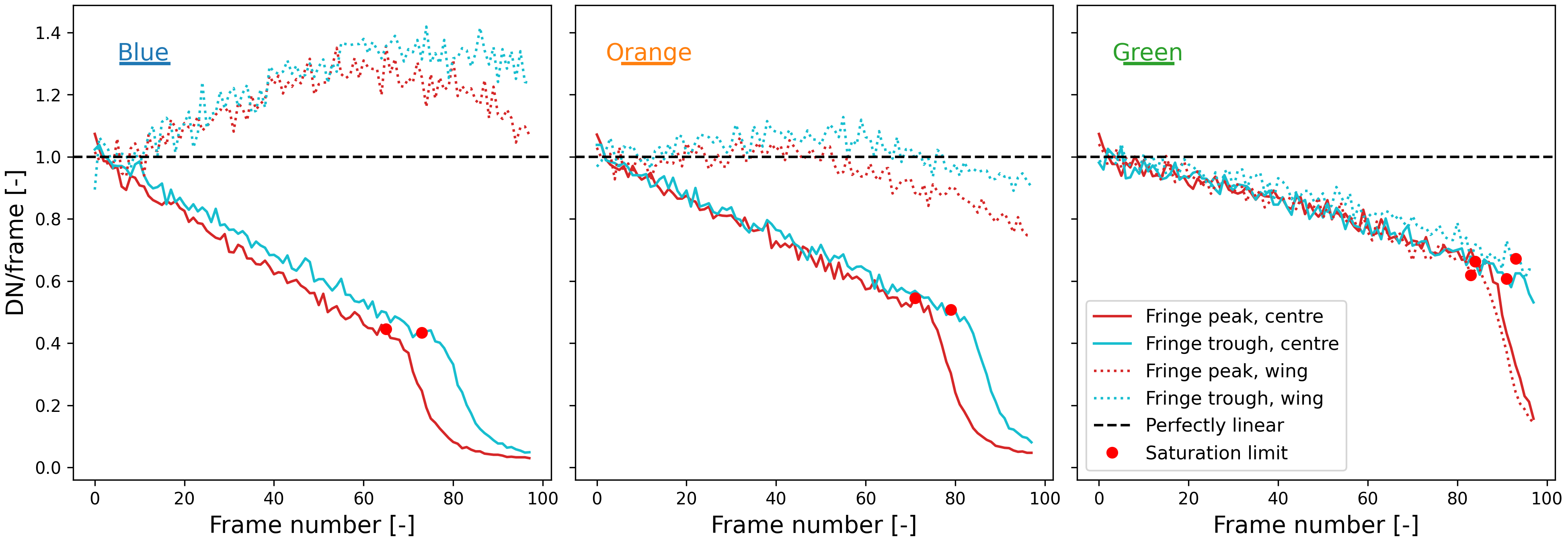}
    \caption{Changing ramp shapes due to fringe peaks and troughs in band 1A. We show the subsequent frame difference, divided by the mean of the first six frames ($\frac{DN_{i+1}-DN_{i}}{\overline{\Delta DN}}$) to normalise the data for visualisation purposes. The first six frames are chosen to still probe an approximately linear part of the ramp. A perfectly linear ramp should then be centred around 1. The stronger the deviation from linearity, (1) the less straight the ramps are, and (2) the quicker they diverge from unity. Cases shown in the panels from left to right correspond to the blue (left), orange (centre), and green (right) ramps in Fig.~\ref{fig:bfe_contrast}, respectively. From left to right, the contrast between the bright central pixel and a neighbour decreases, and the ramps converge to similar shapes. The ramps in the wings for the high-contrast cases show the typical rate-space BFE signature. The point of saturation per ramp is indicated by a red dot on the ramp.}
    \label{fig:fringe_contrast}
\end{figure*}

We recall that the charge in the detector can, and will, migrate as soon as any kind of $DN$-level contrast between neighbouring pixels occurs. This is more challenging to see in the brightest pixels, but shows up in the fainter ramp as increasingly more charge being collected, as previously shown in Fig.~\ref{fig:bfe_contrast}, resulting in the slope (flux) estimated from this ramp being higher than it truly is. Aside from contrast due to the PSF sampling, contrast also occurs due to fringing on the detector. The effects on the ramps are demonstrated in Fig. \ref{fig:fringe_contrast}. For visualisation purposes, the ramps are shown in normalised change per frame $\left(\frac{DN_{i+1}-DN_{i}}{\overline{\Delta DN}}\right)$. For the largest contrast case in the left-most panel, the fringe peak at the centre of the PSF deviates more strongly from linearity than the fringe trough. This is likely due to the fringe trough, which is fainter than the fringe peak, receiving more charge from a brighter pixel. In all cases, this effect becomes strongest at higher $DN$-levels. This suggests that the fringe depth is altered by BFE, resulting in shallower fringes than the reflective properties of the different detector layers might suggest.

The BFE in the spectral direction has two consequences. First, when using the current pipeline single linearity correction for both fringe peaks and troughs, we find a systematic effect on the depth of the fringes depending on the dynamic range covered by the pixels. The ramps in the fringe troughs will accumulate extra charge due to charge migration, causing the ramp shape to deviate from the true non-linearity in case of an isolated pixel. This results in a larger slope in the fringe trough, therefore reducing the fringe depth. In this case the true fringe pattern caused purely by the interference in the spectral direction in the detector is not probed, but rather the result of both interference and migration.

\begin{figure*}[h!]
    \centering
    \includegraphics[width=\textwidth]{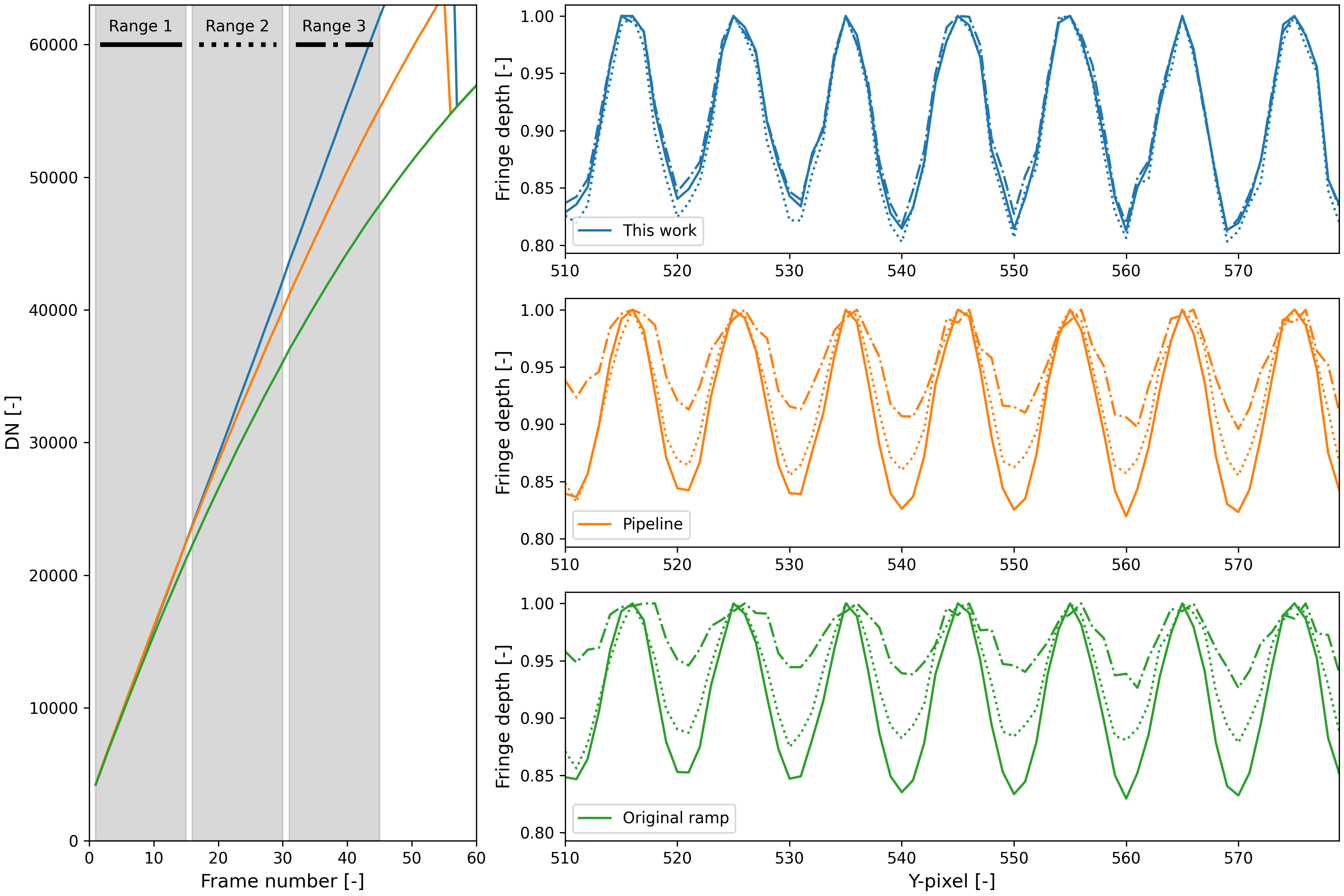}
    \caption{Difference in fringe depth depending on what section of the ramp is used to determine the slope. The right column shows the results from this work (top), the current pipeline (middle), and the original, uncorrected ramp (bottom). The raw ramps are presented in the panel on the left, where the regions used to determine the slope per line style (corresponding to the line styles in the right column) are shown in grey. The fringe depth should be constant if the ramp is linearised.}
    \label{fig:fringes_up_the_ramp}
\end{figure*}

Second, we can do an exercise where we use different parts of the ramp to determine the mean of the frame differences. When the non-linearity in the ramp is not adequately corrected for, the fringe depth will change depending on what part of the ramp is used to determine the rate. As we climb the ramp to higher accumulated $DN$, the effects of both charge migration and non-linearity increase, resulting in shallower fringes the higher the frame number. We note that we do this for demonstration purposes only, to highlight the change in fringe depth with charge spilling. The pipeline uses the full ramp up to saturation to determine the rate, not a sub-section. The true fringe depths per method will therefore be somewhere in the middle of the patterns shown. We demonstrate this effect in Fig.~\ref{fig:fringes_up_the_ramp}, where the ranges used to determine the slopes are shown in the left panel. The resulting fringe patterns are shown on the right. The uncorrected ramp, plotted in the bottom panel, clearly shows the previously described effect: the fringes in range 1 where the BFE impact is the smallest are significantly deeper than the fringes in range 3, and there even seems to be a slight phase shift, which implies that the static fringe flats used by the pipeline will be (1) modifying the frequency of the fringes in a systematic way, and (2) subsequently impacting the residual fringe correction downstream. This is already improved in the middle panel, where the linearity correction derived from extended illumination that is part of the current pipeline is displayed. There is still a significant difference between the curves, translating to a systematic fringe residual. However, if the ramp is truly linearised, there should be no change in the fringe depth, since this is a purely optical interferometric effect. In the linearity correction per pixel derived in this work shown in the top panel, the effect is mitigated and the fringe depth is independent of the $DN$-range used for determining the rate. When calculating the difference between the fringing at the start of the ramp and the end, the standard deviation on this value is improved by $\sim$60\% when comparing the standard JWST pipeline output to the work here.

\subsection{Interpolating fit coefficients}
\label{sec:interp}

\begin{figure*}
 \includegraphics[width=.48\linewidth]{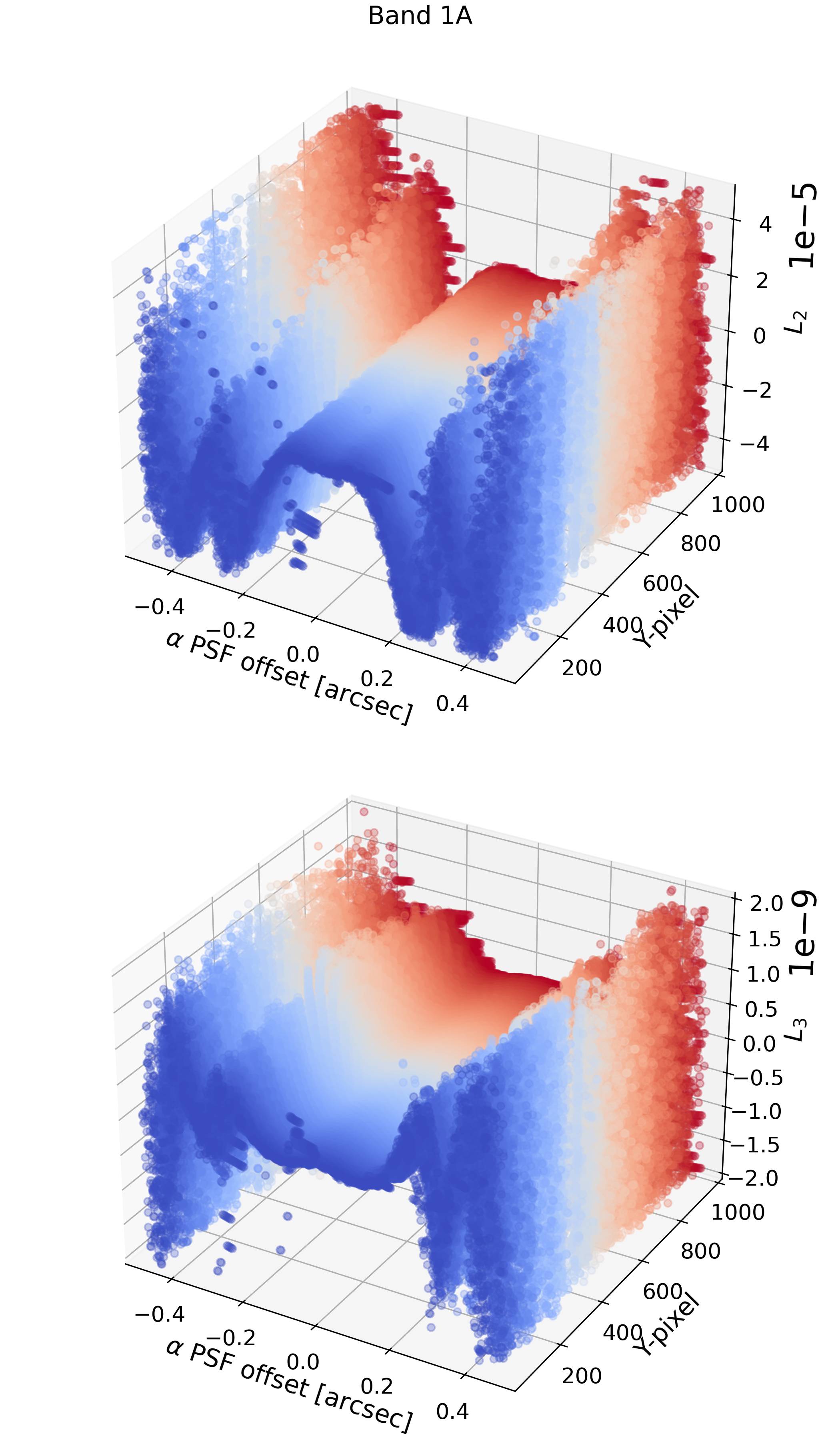} \hfill
 \includegraphics[width=.48\linewidth]{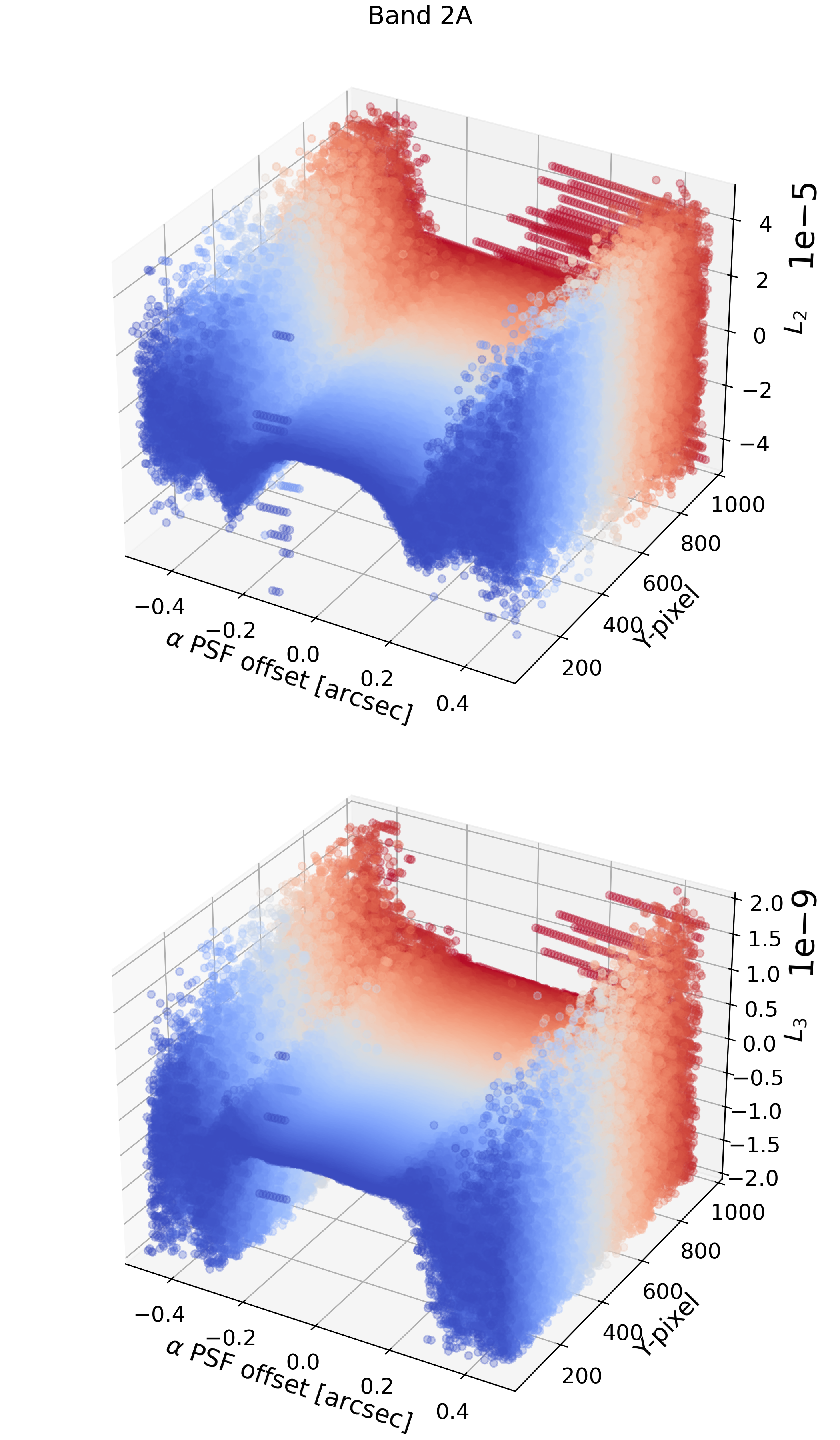}
 \caption{Change in polynomial coefficients $L_2$ and $L_3$ with offset from PSF peak in bands 1A (left) and 2A (right).}
   \label{fig:12a_polyfits}
\end{figure*}

One way to apply the new linearity corrections, is to find the reference solution where the centroid is as close as possible to the science observation, and apply the nearest solution, which we refer to as the grid solution. However, some observations might fall outside the grid of observations performed here by PID~3779. The linearity solution per pixel should only depend on the illumination; that is, the contrast between the individual pixels regardless of where those pixels are located on the detector. To test whether we can apply the solution derived here to observations with a larger offset from the grid, we plot the second and third order coefficients ($L_2$ and $L_3$, recall that $L_0=0$ and $L_1=1$) as a function of offset from the PSF centre and Y-pixel (detector row, sampling the MRS spectral direction) probed in order to take changes due to fringing into account. The offsets are determined by finding the location of the PSF peak per observation (which may not coincide exactly with a pixel) using methods similar to \citet{ref:24PaArLa}, and relating the specific pixel coordinate to how far this is from the PSF peak coordinate in $\alpha$. We assume that the influence of $\beta$ is only in the relative flux per slice, and does not change the way the PSF is sampled per pixel.

The results of this analysis are presented in the left column of Fig.~\ref{fig:12a_polyfits} for band 1A. Quite a clear trend for both coefficients becomes visible with both detector row (Y-pixel), and part of the PSF probed. Near the wings of the PSF the data becomes noisy, and parts of this have been cut from the plot by limiting the range from -0.41 to 0.41~arcsec (recall that the pixel resolution is $\sim$0.2~arcsec), which is wider than the FWHM of the PSF, indicating that the majority of the PSF can be modelled in this way. The results for the other sub-bands of channels 1 and 2 can be found in App.~\ref{app:poly_fits}, which are very similar to the trends in Fig.~\ref{fig:12a_polyfits}. In fact, the main difference is the general broadening and flattening of the trends. To demonstrate this, band 2A can be found in the right column of Fig.~\ref{fig:12a_polyfits} (where the angular size of the pixels is the same as in 1A), which shows this effect clearly. This is likely caused by the gradual increase in PSF size as the wavelengths probed increase due to diffraction, causing the relative contrast between neighbouring pixels to decrease. When the contrast is smaller, the ramps are less affected by charge migration and the coefficients will be more consistent between pixels, since this is closer to the ideal case of uniform or flat illumination.

\begin{figure}[h!]
	\centering
    \includegraphics[width=0.9\columnwidth]{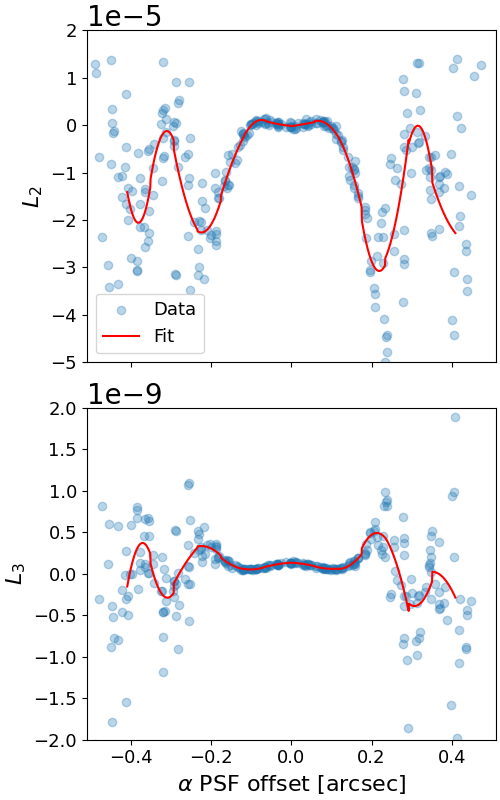}
    \caption{Example of one slice of $L_2$ and $L_3$ fitted with splines in band 1A.}
    \label{fig:fit_demo}
\end{figure}

\begin{table}[t]
\caption{Fit binning and padding parameters used per channel to fit the trends in $L_2$ and $L_3$.}
\label{tab:fit_param}
\centering
\begin{tabular}{lll}
\hline \hline
\textbf{Channel} & \textbf{\# bins} & \textbf{Padding ($\mathbf{\Delta \alpha}$ [arcsec])} \\ \hline
1                & 15               & 0.02                               \\
2                & 8                & 0.04      \\ \hline                        
\end{tabular}
\end{table}

In order to create a noiseless grid of coefficients, we fit the trends on each detector row (Y-pixel). By going row-by-row, small changes in the coefficients per row due to fringing will be folded into the solution as well. This is also dependent on how the PSF is sampled, and we make sure to include this effect by not collapsing multiple detector rows. The trend to fit, especially in the shortest wavelengths, contains steep increases and sharp edges. In order to incorporate these, the data is split into chunks of $\alpha$-offsets, and a series of quadratic splines is fit through these chunks. When retrieving the $L_2$ and $L_3$ for a specific pixel, the spline fit to the chunk within which its $\Delta \alpha$ falls is used to interpolate the coefficients. However, we do include a padding factor either side of the partial fits, where data points from other sections are included in the fitting procedure. This is done to create a smoother function through the data points. The number of bins used, and the padding in $\alpha$-offset, are included in Table~\ref{tab:fit_param}. This process is repeated for all Y-pixel rows individually. One slice of these fits is shown in Fig.~\ref{fig:fit_demo}, where the top figure corresponds to $L_2$ and the bottom plot to $L_3$. The longer the wavelengths probed, the easier it becomes to generate a smooth fit, due to the lack of sharp edges.

In Fig.~\ref{fig:linearised_ramps} the qualitative results of linearing the ramps using the interpolated solution are also shown. Overall, the interpolated solution agrees well with the grid solution, and both perform better than the current pipeline. However, due to limited S/N past $\Delta \alpha\approx 0.2$~arcsec, the interpolated solution does not always perfectly match the grid solution. Despite these limitations, 70\% of the rates derived from the interpolated correction are within 10\% of the rate derived from the grid correction, which is still a significant improvement compared to the current pipeline.

Next, the noise-free coefficients are tested on the same grid samples and test observation as in Sect.~\ref{sec:stats}, to make sure the ramps also become linear when using the interpolated values. In order to find the best coefficients per ramp, and therefore per pixel, a few steps must be performed prior to running the pipeline with the new coefficients. In order to have a detector slope image from which the PSF centroid can be determined, firstly the rates of the raw files must be calculated. This can be done by simply running the first stage (\texttt{CALDETECTOR1}) of the current pipeline. Secondly, the centroid in $(\alpha,\beta)$ per observation must be found on the detector \citep{ref:24PaArLa}. This is then the centre of the PSF, and the specific $\alpha$-coordinate of each pixel will give the offset from the PSF peak in arcseconds. From the fits per detector row, for each band, the coefficient per pixel can then be extracted. This is saved per pixel, and compiled into a reference file similar to the ones generated directly from the grid and the current pipeline files. This can then be fed into the \texttt{linearity} step of the current pipeline. We note that the edges of the fits are not used due to the increase in noise, and these parts are simply replaced by the values in the solution of the standard calibration pipeline. This is past a $\Delta \alpha$ of $\pm$0.4. The linearity of the resulting data is assessed in the same way as in Sect.~\ref{sec:stats}. The statistics of the ramps can be found in the right-most columns of Figs.~\ref{fig:stats_self} and \ref{fig:stats_ref}. In Fig.~\ref{fig:stats_self}, the interpolated method does not linearise the ramps as well as directly applying the grid, which is to be expected since the former case is self-calibrating the data, the ramps are still much more centred around linearity than the current pipeline (uniform) solution. However, in the case of the test observation presented in Fig.~\ref{fig:stats_ref}, the interpolation method works much better than directly applying the grid, since the observation was outside of the grid points. The improvement in spread on the error is up to 70\%. This demonstrates that the way the PSF is sampled by the pixels is more important when correcting the ramps, than its exact location on the detector. The interpolation method therefore shows a lot of promise, and could even be applied to targets that do not use target acquisition.

\subsection{PSF FWHM}
The BFE, when uncorrected, can broaden the PSF and line-spread function (LSF) \citep{ref:23ArLaRi}. The specifics of this will depend on the sub-pixel pointing of the telescope, and will not only affect the estimated flux, but also the accuracy of PSF subtractions for high-contrast imaging applications as discussed in Sect.~\ref{sec:introduction}. So far reference stars of comparable brightness have been used for PSF subtraction to uncover fainter companions, either by using principal component analysis (PCA) \citep{ref:24CuPaBa}, or direct subtraction, which have worked quite well. However, to truly push the contrast to its limits, PSF subtraction must be performed on the detector itself, rather than after building the rectified spectral cube \citep{ref:23LaMoAr}. In order to generate accurate PSF models with sub-pixel pointing, the first step is to remove the BFE signature from the ramps for each dither position as we have done in this work. This should mitigate the effect of BFE on the intrinsic FWHM of the PSF.

\begin{figure*}
	\centering
    \includegraphics[width=0.9\textwidth]{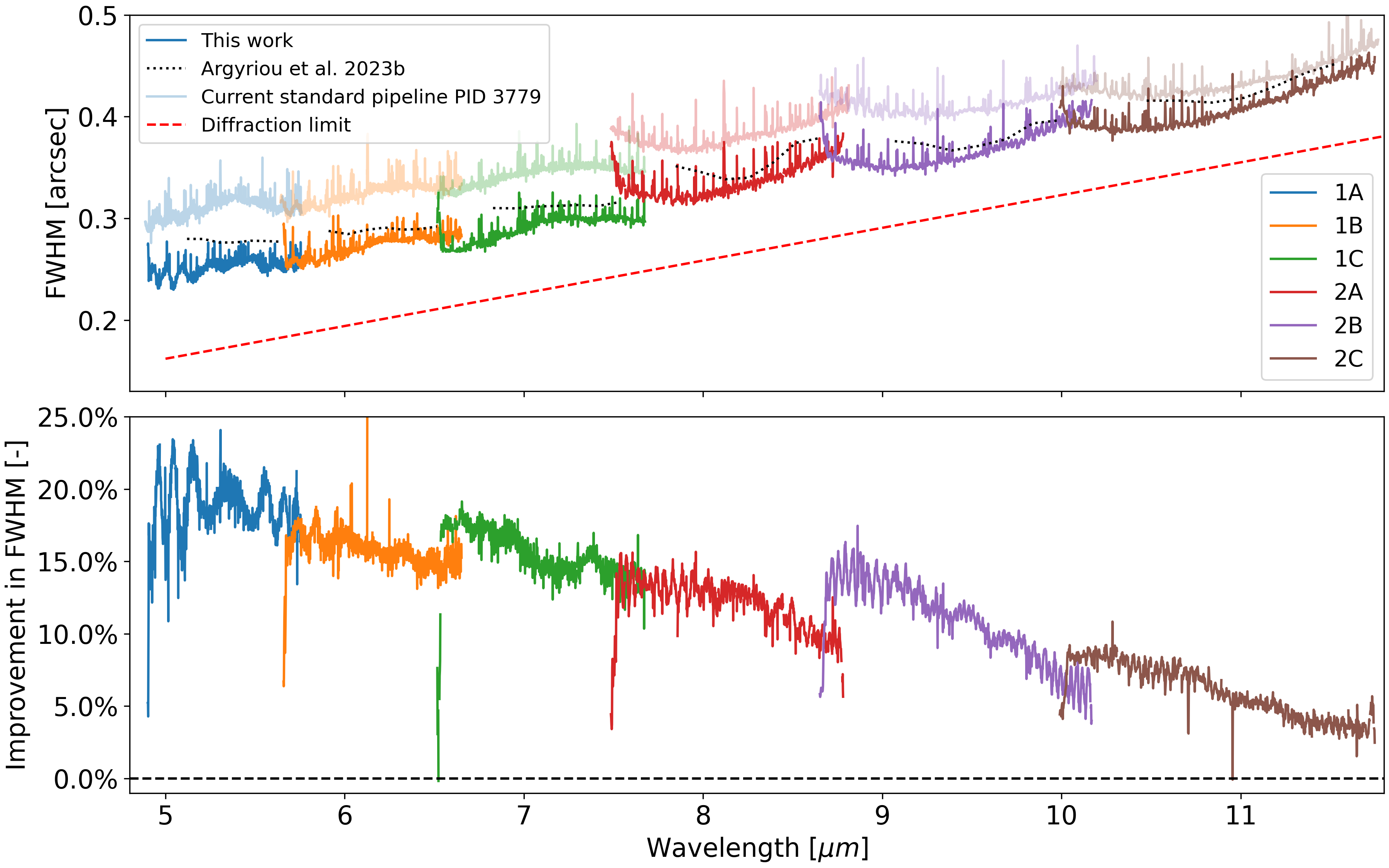}
    \caption{Comparison of detector FWHM as measured in \citet{ref:23ArGlLa} (black dotted lines), from this data set but processed with the current pipeline (lower opacity colours), and after processing using the optimised corrections (full-opacity colours). The top shows the FWHM values, and the bottom the percent-difference in FWHM between the current pipeline and this work.}
    \label{fig:fwhm}
\end{figure*}

In the top panel of Fig.~\ref{fig:fwhm} a comparison is shown between the FWHM on the short wavelength detector after commissioning from \citet{ref:23ArGlLa}, and the FWHM derived from 10-Lac for both the current pipeline and this work. Previously, multiple rows on the detector had to be combined in order to properly sample the detector PSF (see \citealt{ref:23ArGlLa}). With a total of 72 dithers of 10-Lac at our disposal, there is no longer a need to collapse multiple pixel rows, and the PSF is sampled well for each individual row. We determine the FWHM similarly to how it was done during commissioning \citep{ref:23ArGlLa}. The data is normalised by summing all the flux in a detector row and dividing by this value. Subsequently, the PSF in the brightest slice from all the different dithers are centred on top of each other. The FWHM is then determined by fitting a Gaussian, and calculating the FWHM of this Gaussian.

The new solution yields slightly better results than \citet{ref:23ArGlLa}. The data used shortly after commissioning is relatively faint (low accumulated $DN$), and therefore not significantly affected by BFE. However, when measuring the FWHM of 10-Lac after processing the data with the standard calibration pipeline, a broadening can be observed. This broadening is quantified in the bottom panel of Fig.~\ref{fig:fwhm}, where the percent-difference between the old and new solutions are shown. For bright sources, this work results in a PSF up to 20\% narrower than the current calibration pipeline, which is in line with the broadening found in the MIRI Imager \citep{ref:23ArLaRi}. This improvement decays as longer wavelengths are probed. There are two potential explanations for this. First, it could be another dynamic range effect, where the star does not probe enough of the ramp in longer wavelengths, and the effect of BFE is lessened. Second, the PSF broadens for longer wavelengths (we already observed the effect on the ramp shapes in Sect.~\ref{sec:interp}) potentially resulting  the effect of BFE being less severe. It could also be a combination of both.

Prior to commissioning, the MIRI/MRS PSF FWHM, which by design should have been diffraction limit in the full instrument wavelength range, had been observed to not be diffraction limited, with usable data only up to 8~microns. During commissioning it was found that the PSF does not reach the diffraction limit in the longer wavelengths either, the main cause thought to be the scattering within the detectors \citep{ref:23ArGlLa}. This indeed seems to be true, although the scattering contribution may be slightly less than initially thought. The data used to measure the FWHM during commissioning was likely not bright enough to be significantly affected by BFE. Even after correcting for it now, the diffraction limit is not reached. However, this does clearly demonstrate the effect of dynamic range on the PSF broadening, and our ability to correct the effect. Furthermore, since the PSF is narrower than the one derived from commissioning data (which did not saturate), it shows that BFE starts prior to saturation (as also clearly visible in the left panel of Fig.~\ref{fig:fringe_contrast}), and cutting the neighbouring ramps after saturation of the central ramp still results in a noticeable broadening (which is the approach that can currently be applied in the JWST pipeline, but is not turned on by default for MIRI specifically, at the time of writing). More details on a detector-based PSF model, detector-based PSF-weighted extraction and PSF subtraction that includes the corrections derived will be available in Paper III in this series.

\section{Discussion}
\label{sec:discussion}

\subsection{Sinusoid residual}

\begin{figure*}[]
    \centering
    \includegraphics[width=\textwidth]{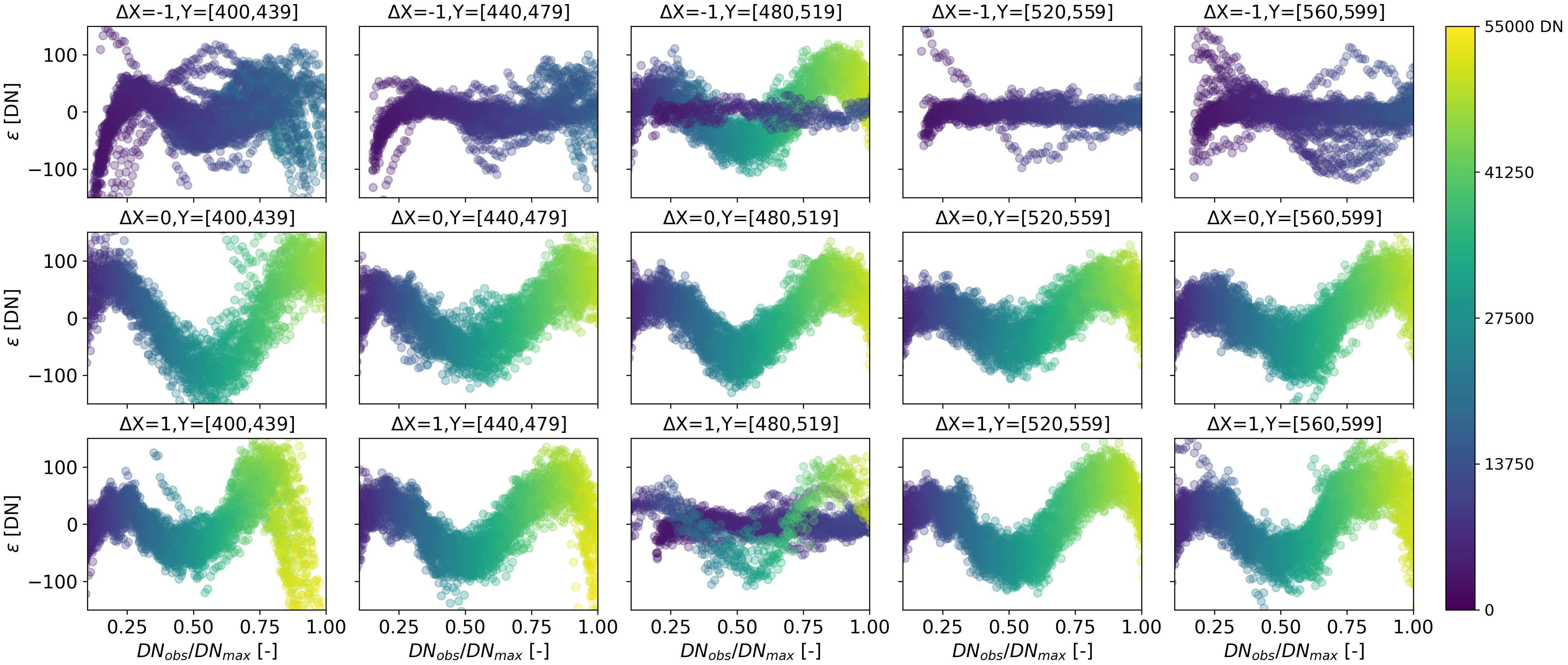}
    \caption{Residuals of linearised ramp after subtracting a linear fit. The X-offset is with respect to the brightest central pixel trace, while the Y-coordinates are a range. The x-axes are the observed $DN$ of the original ramp, divided by the maximum $DN$-value reached. The colour map shows the $DN$-level of the specific time-sample indicated by the dot, where 55000~$DN$ roughly corresponds to the imposed saturation limit (55030~$DN$). The residuals show a systematic sinusoid signature with $DN$-level reached, around a noise floor of approximately 50~$DN$.}
    \label{fig:sinusoid}
\end{figure*}

If one were to subtract the third order polynomial from the raw ramps, a systematic remains in the residuals. Although it has an amplitude of $\pm$100~DN ($\pm$50 when ignoring the spread in noise), which is only 0.18\% of saturation, it does not have the same characteristics as random, Gaussian noise. The residuals over the ramp are plotted in Fig.~\ref{fig:sinusoid}, where ramps that probe the full dynamic range show a distinct sinusoidal pattern, with two positive peaks within the dynamic range of the ramp. While it may seem that ramps that are not as bright do not contain this residual, it could be that the first part of the sinusoid (up to $\sim 25\%$ of the saturation limit) which is approximately linear, is included in the linear ramp fit and would therefore not show up in the residuals. The shape of the residuals is not in line with the expectations from the classical non-linearity and BFE; it falls outside the periodicity of telescope jitter ($\sim$0.04~Hz, \citealt{ref:22jitter}), and we have observed the same effect in the MIRI Imager. Specifically, in PID~1464, observation 5, which is shown in Fig.~5 of \citet{ref:23ArLaRi} to reach a similar dynamic range as the data presented here. However, there are two other factors that influence the ramp shape: the reset effect and saturation.

The first is the charge that is not properly flushed out after a reset, which shows up as reset decay in the ramp (see also \citealt{ref:23MoDiAr}). At the start of the exposure, the rate of the ramp will be higher than it should be, and this slowly decays over 100 or more FASTR1 groups, equivalent to 275.5~s. This effect is proportional to the incoming flux before the pixels start integrating. Therefore, the sharp hook at the start of the sinusoid is likely due to the reset effect settling down. 

On the other hand, when nearing saturation the ramps curve strongly towards a plateau (as seen in the left panel of Fig.~\ref{fig:bfe_contrast}). The hook at the end of the sinusoid can likely be attributed to this part. 

While the ramp profile near saturation may be similar per observation, the reset effect is not. The latter can depend on how much charge is on the detector, the pixel indium bump bonds, or at the input to the (pixel) unit cell buffer amplifier before the exposure starts \citep{ref:23ArLaRi}. This may vary from observation to observation, depending on what was targeted prior to the science observation. Since it is not a constant effect, a constant correction derived from this particular data set will not be universal for all observations. Therefore, while it seems the sinusoidal signature is something that can be included in the ramp fits, we do not do so.

If the sign of the sinusoid pattern is indeed constant with $DN$-level reached, it will cause a slight change in slope depending on how far the samples reach up the ramp. This means that depending on how bright the observation is in a single integration, the flux derived from this will be slightly biased. When only probing the first $\sim 15000$~DN of the ramp, the derived rate will be higher than it should be. On the other hand, when an observation is probing nearer to saturation, the effects might even out slightly. Additionally, the spectrophotometric calibration is based on 10-Lac data from PID~1524 (PI: D. Law), where 20~groups were probed (corresponding to range 1 in Fig.~\ref{fig:fringes_up_the_ramp}), meaning that it may inherently correct for some of the reset when these exposures do not probe the dynamic range very well, which would conversely result in inaccurate flux for exposures that do.

%\subsection{IPC versus BFE}
%\label{sec:ipc}
%In Sect.~\ref{sec:data_reduction} we mentioned that the IPC correction step is applied prior to deriving the linearity correction, as this step affects the ramp shape. IPC is electronic crosstalk between neighbouring pixels resulting in the pixels becoming correlated with each other. Crosstalk was measured to be around the 3\%-level in the MIRI Imager prior to launch, most of which was attributed to IPC \citep{ref:15RiReMo}. These measurements were repeated in-flight, yielding similar results \citep{ref:23EnRe}. By finding the correlation between a central pixel and its neighbours, a correlation matrix is found, which is converted to a kernel used to deconvolve the raw signal, and remove the IPC signature from the ramps. Both BFE and IPC cause \textit{DN}s to be dumped into neighbouring pixels; the former by physically migrating charge, the latter due to an electronic effect. For the MRS deriving the IPC using the internal calibration source measurements is a degenerate process with BFE due to the significant illumination contrast imposed by the spectral fringes. This is still work in progress.

\subsection{Linearity solutions at longer wavelengths}
\label{sec:long_wvl_sol}
Due to the Spectral Energy Distribution (SED) of 10-Lac dropping off at longer wavelengths (channels 3 and 4), it becomes impossible to derive a linearity solution at these wavelengths. While the contrast between pixels will be less due to the broadening of the PSF, therefore resulting in a lower variation in ramp shapes, it would still be beneficial to the quality of the data to be able to apply the same methods. Ideally, this would be a target with few spectral features, and allow for stable pointing (targets moving up to 75~mas/s\footnote{\url{https://jwst-docs.stsci.edu/methods-and-roadmaps/jwst-moving-target-observations/jwst-moving-target-proposal-planning/moving-target-acquisition-and-tracking}}). Recently, an empirical spectrophotometric calibration and fringe correction similar to \citet{ref:23GaArSl} has successfully been derived in longer wavelengths based on asteroids, though this was done after building the rectified spectral cubes in stage 3 of the pipeline, rather than on the detector \citep{ref:23PoSaBa}. Due to their red SED, these asteroids would also be suitable for deriving linearity solutions at long wavelengths.
%, if sufficiently long exposures with enough stability can be taken. \citet{ref:23PoSaBa} give the accuracy of the pointing for their target. Especially in channel 4 the offsets can be as large as 0.8 spaxels, amounting to approximately 0.28~arcsec \citep{ref:23LaMoAr}, which is not stable enough for the methods described here. However, the measurement of an offset will be much less precise in channel 4. Furthermore, when focusing solely on BFE calibration, our results in Sect.~\ref{sec:interp} suggest that interpolating fit coefficient works. Whether the asteroid is stable enough is an open question, but 
The methods described here might also be possible for a different red unresolved target such as a bright planetary nebula.

\section{Conclusion}
\label{sec:conclusion}
In this work, we argue for the need for taking the contrast between pixels into account when deriving linearity corrections, and demonstrate the level of improvement that can be achieved. The tailored data from the Cycle 2 calibration programme PID~3779 provide a consistent dataset, and with the impressive pointing accuracy of JWST we are able to model trends in the ramps. In terms of self-calibration on the PID~3779 data themselves, this results in a 70-90\% smaller spread on the distribution of residual deviation from linearity. Furthermore, the ramp shapes seem to largely depend on what part of the PSF is sampled. By plotting the polynomial coefficients as a function of offset from the PSF centre, it becomes clear that this can be a spline-based calibration solution that can be interpolated. In the one case that had the correct set-up to test this, the resulting improvement on the spread is up to 70\%. In general, the improvement compared to the current pipeline (which assumes perfectly flat illumination) will vary depending on the location of the source with respect to the grid, as well as how much of the pixel dynamic range the pixels cover. When removing the BFE signature, the fringes become deeper and more consistent up-the-ramp, as is expected from the detector physics. Similarly, in the spatial direction, the FWHM for sources covering the pixel dynamic range is reduced by up to 20\%. This is extremely important for science cases such as high-contrast imaging with the MRS, and measuring accurate fluxes. In Paper II we will quantify the change in the derived integrated-point source $DN$/s count rate. The improvement in PSF FWHM indicates that simply removing the frames of the neighbouring pixels after the brightest pixel saturates is not enough. Rather than only migrating once a pixel saturates, the charge already migrates before this point.

These improvements can currently only be derived for the shortest wavelengths. In the longer wavelengths, channels 3 and 4, 10-Lac becomes too faint, which was already known at the time of writing the proposal (see Fig.~\ref{fig:range_per_band}). Due to this, we do not probe the dynamic range required to derive a linearity correction in these wavelengths. Ideally this work should be repeated using a target that is brighter in the longest wavelengths, such as an asteroid, to reproduce the three-by-three intra-pixel mosaic grid presented here. Based on the polynomial coefficient trends observed in channels 1 and 2, it does become clear that the difference in ramp shapes across the PSF reduces with increasing wavelength. This is either due to diffraction expanding the PSF and reducing the illumination contrast, or the reduced dynamic range in these regions, but the leading cause is currently not known. The question therefore remains to what degree the much larger PSF in the longer wavelengths reduces the contrast between pixels. While the spatial differences are most significant in the shortest wavelengths, spectral variation due to fringing will remain a large component, even at the longer wavelengths.

Now that one of the major detector effects has been addressed, we will use the self-calibrated 10-Lac data to address the next set of goals based on our bottom-up approach. In Paper~II, the data will first be used to characterise the changes in fringing with PSF sampling, as well as characterising systematics in the spectrophotometric solution. In Paper~III, we will derive the master PSF on the detector, for the purpose of extracting spectra directly from the detector, and subtracting the PSF on the detector level for high-contrast imaging applications. Finally, in Paper~IV the identified trends will be used to address these issues for (semi-)extended illumination.

\begin{acknowledgements}

This work is based on observations made with the NASA/ESA/CSA James Webb Space Telescope. The data were obtained from the Mikulski Archive for Space Telescopes at the Space Telescope Science Institute, which is operated by the Association of Universities for Research in Astronomy, Inc., under NASA contract NAS 5-03127 for JWST. These observations are associated with programs \#3779 and \#1050. \\

Danny Gasman, Ioannis Argyriou, thank the European Space Agency (ESA) and the Belgian Federal Science Policy Office (BELSPO) for their support in the framework of the PRODEX Programme. \\
% Patrick Kavanagh thanks the European Space Agency (ESA) and Enterprise Ireland for their support in the framework of the PRODEX Programme. \\

P.J.K. acknowledges financial support from the Science Foundation Ireland/Irish Research Council Pathway programme under Grant Number 21/PATH-S/9360. \\

The work presented is the effort of the entire MIRI team and the enthusiasm within the MIRI partnership is a significant factor in its success. MIRI draws on the scientific and technical expertise of the following organisations: Ames Research Center, USA; Airbus Defence and Space, UK; CEA-Irfu, Saclay, France; Centre Spatial de Liége, Belgium; Consejo Superior de Investigaciones Científicas, Spain; Carl Zeiss Optronics, Germany; Chalmers University of Technology, Sweden; Danish Space Research Institute, Denmark; Dublin Institute for Advanced Studies, Ireland; European Space Agency, Netherlands; ETCA, Belgium; ETH Zurich, Switzerland; Goddard Space Flight Center, USA; Institute d'Astrophysique Spatiale, France; Instituto Nacional de Técnica Aeroespacial, Spain; Institute for Astronomy, Edinburgh, UK; Jet Propulsion Laboratory, USA; Laboratoire d'Astrophysique de Marseille (LAM), France; Leiden University, Netherlands; Lockheed Advanced Technology Center (USA); NOVA Opt-IR group at Dwingeloo, Netherlands; Northrop Grumman, USA; Max-Planck Institut für Astronomie (MPIA), Heidelberg, Germany; Laboratoire d’Etudes Spatiales et d'Instrumentation en Astrophysique (LESIA), France; Paul Scherrer Institut, Switzerland; Raytheon Vision Systems, USA; RUAG Aerospace, Switzerland; Rutherford Appleton Laboratory (RAL Space), UK; Space Telescope Science Institute, USA; Toegepast- Natuurwetenschappelijk Onderzoek (TNO-TPD), Netherlands; UK Astronomy Technology Centre, UK; University College London, UK; University of Amsterdam, Netherlands; University of Arizona, USA; University of Bern, Switzerland; University of Cardiff, UK; University of Cologne, Germany; University of Ghent; University of Groningen, Netherlands; University of Leicester, UK; University of Leuven, Belgium; University of Stockholm, Sweden; Utah State University, USA. A portion of this work was carried out at the Jet Propulsion Laboratory, California Institute of Technology, under a contract with the National Aeronautics and Space Administration.\\

We would like to thank the following National and International Funding Agencies for their support of the MIRI development: NASA; ESA; Belgian Science Policy Office; Centre Nationale D'Etudes Spatiales (CNES); Danish National Space Centre; Deutsches Zentrum fur Luft-und Raumfahrt (DLR); Enterprise Ireland; Ministerio De Economiá y Competividad; Netherlands Research School for Astronomy (NOVA); Netherlands Organisation for Scientific Research (NWO); Science and Technology Facilities Council; Swiss Space Office; Swedish National Space Board; UK Space Agency.\\

We take this opportunity to thank the ESA \textit{JWST} Project team and the NASA Goddard ISIM team for their capable technical support in the development of MIRI, its delivery and successful integration.

\end{acknowledgements}

% References
\bibliography{bib} % bibliography data in aanda.bib
\bibliographystyle{aa} % makes bibtex use aa.bst

\appendix

\section{Observed dither positions}
\begin{figure}[]
    \centering
    \includegraphics[width=\columnwidth]{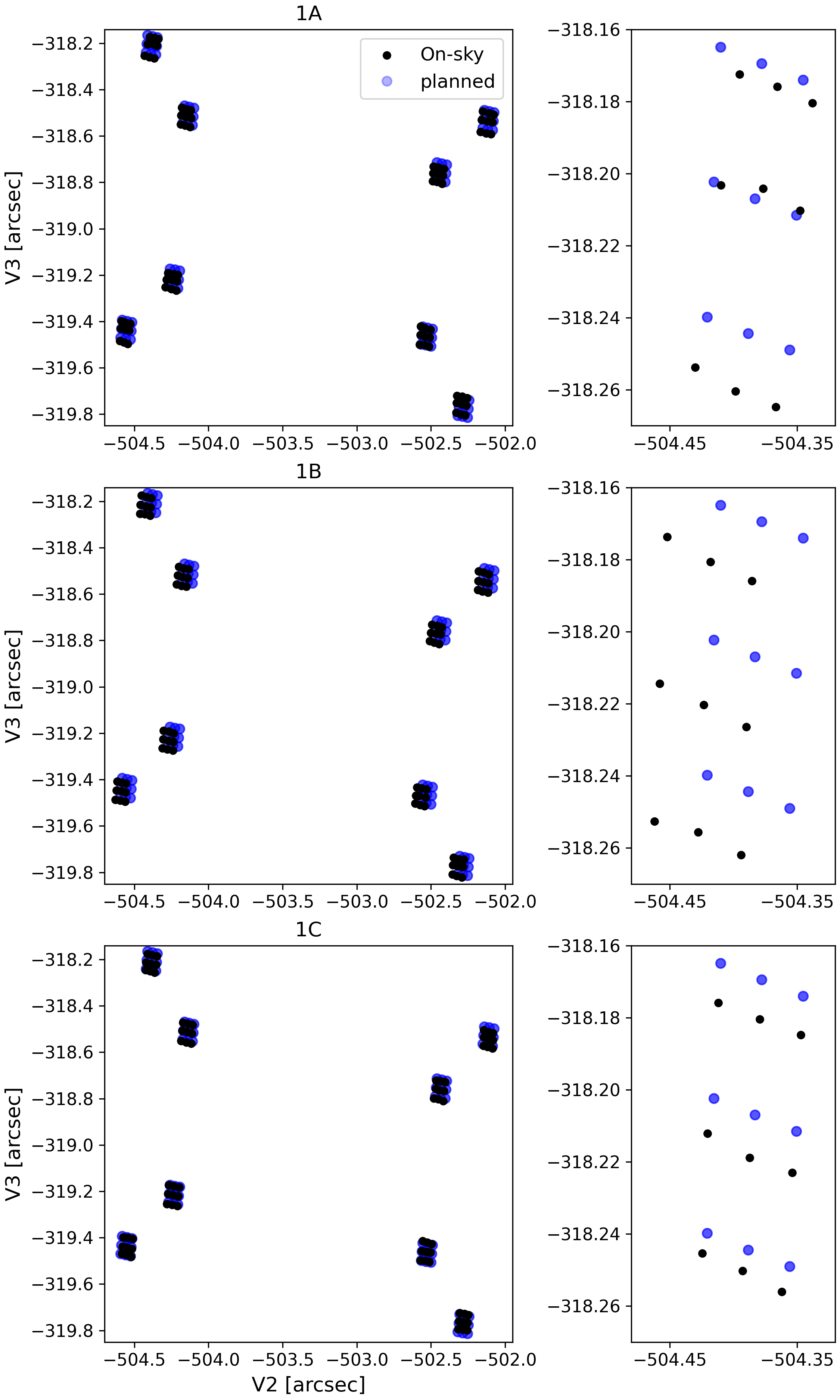}
    \caption{Comparison between requested mosaic (blue) and where the actual observations pointed on-sky (black).}
    \label{fig:observed dithers}
\end{figure}

Fig.~\ref{fig:observed dithers} presents the comparison between planned and real on-sky positions as measured in the data. We note that the relative positions between dithers and within the mosaic is extremely accurate, but the least accurate in the first band (indicated by 1A, top row). This could be a limitation of the centroiding algorithm used, which fits a two-dimensional Gaussian to the detector measurements, and will be least accurate in the extremely under-sampled shortest wavelength band; or the MRS distortion solution. Determining the origin of the uncertainty is beyond the scope of this work.

\section{Dynamic range of ramps per band}
\begin{figure*}[]
    \centering
    \includegraphics[width=0.7\textwidth]{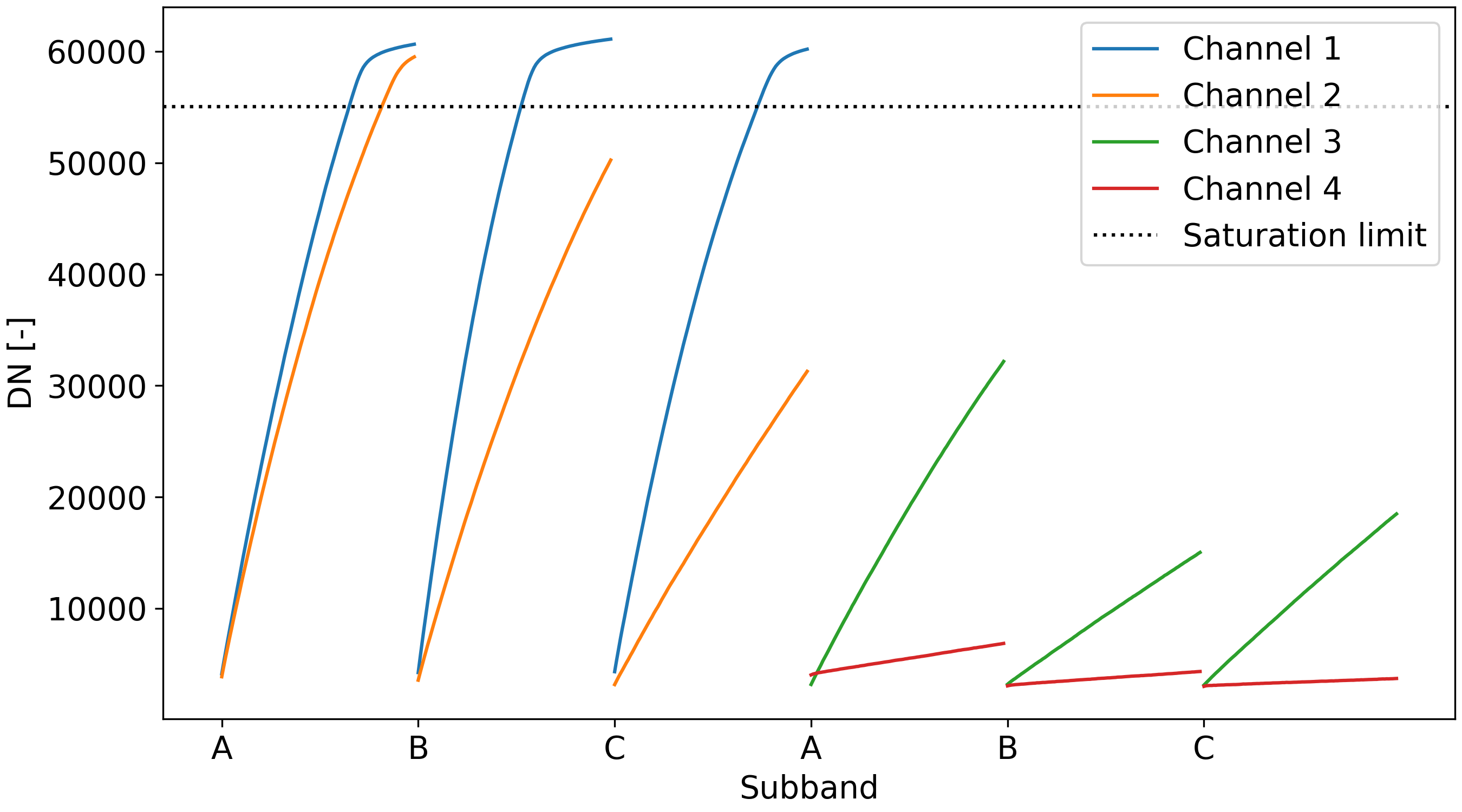}
    \caption{Dynamic range reached per band.}
    \label{fig:range_per_band}
\end{figure*}

The dynamic range of the ramps depends on the $DN$-level reached while probing the ramp with enough samples. In this programme, each ramp consists of 100 samples (number of groups), and saturation is reached in the shortest wavelengths. The dynamic range per band is shown in Fig.~\ref{fig:range_per_band}.

\section{Noise per maximum polynomial fit order}
\label{sec:noise}
\begin{figure}[]
    \centering
    \includegraphics[width=1\columnwidth]{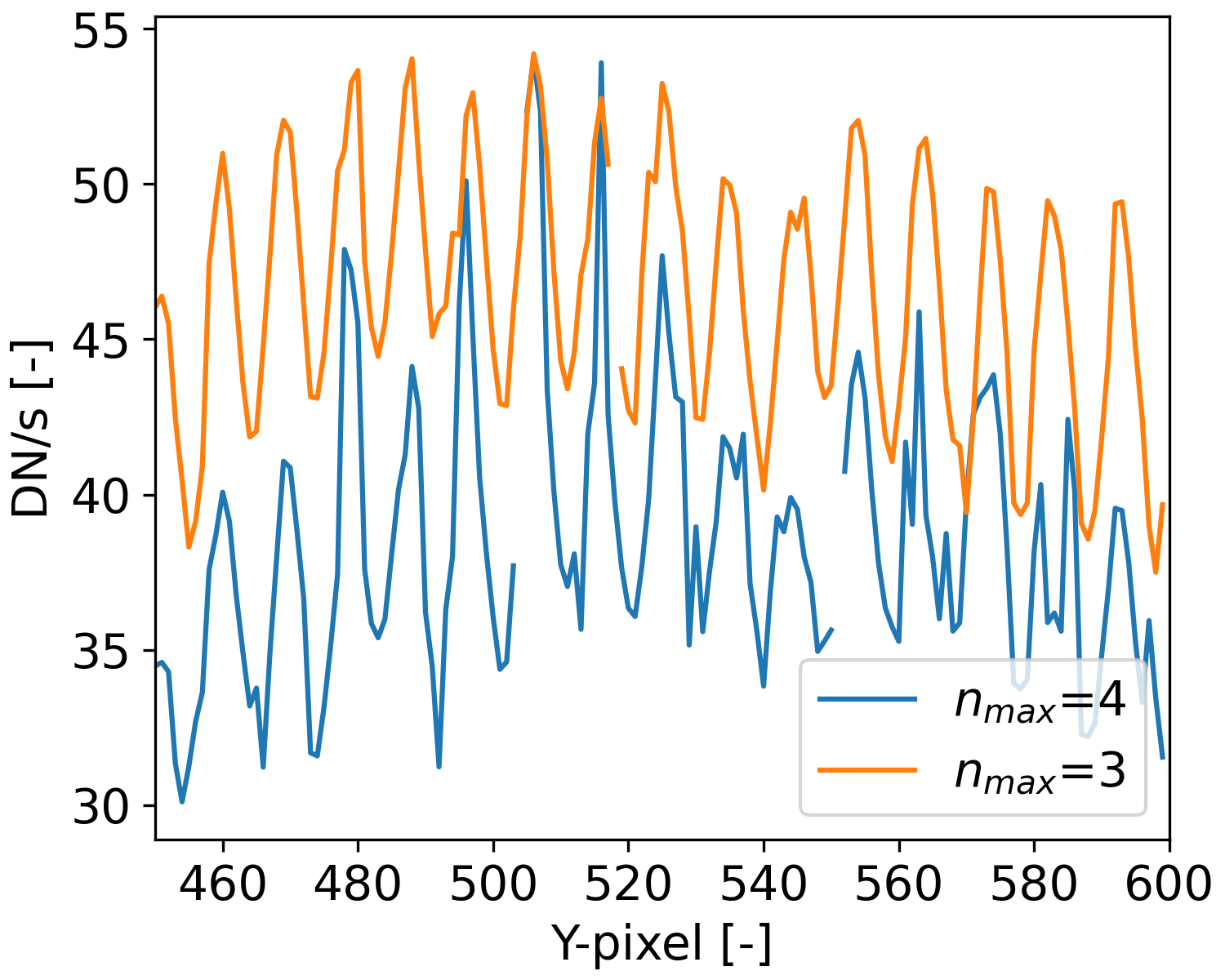}
    \caption{Part of Y-pixel column after running stage~1 of the pipeline, with either the optimal polynomial degree (plateau of Bayesian evidence from Fig.~\ref{fig:bayes_peaks}, $n_{max}=4$), or third order for all ramps ($n_{max}=3$).}
    \label{fig:noise}
\end{figure}

In Fig.~\ref{fig:noise} we demonstrate the increase in noise when using the optimal polynomial orders derived from Bayesian statistics in Sect.~\ref{sec:bayes}, compared to using a third order polynomial for all ramps.

\section{Polynomial coefficient trends in all short detector bands}
\label{app:poly_fits}

The point clouds of the correction coefficients per sub-band, aside form sub-band A, which is shown in the main text.

\begin{figure*}
 \includegraphics[width=.48\linewidth]{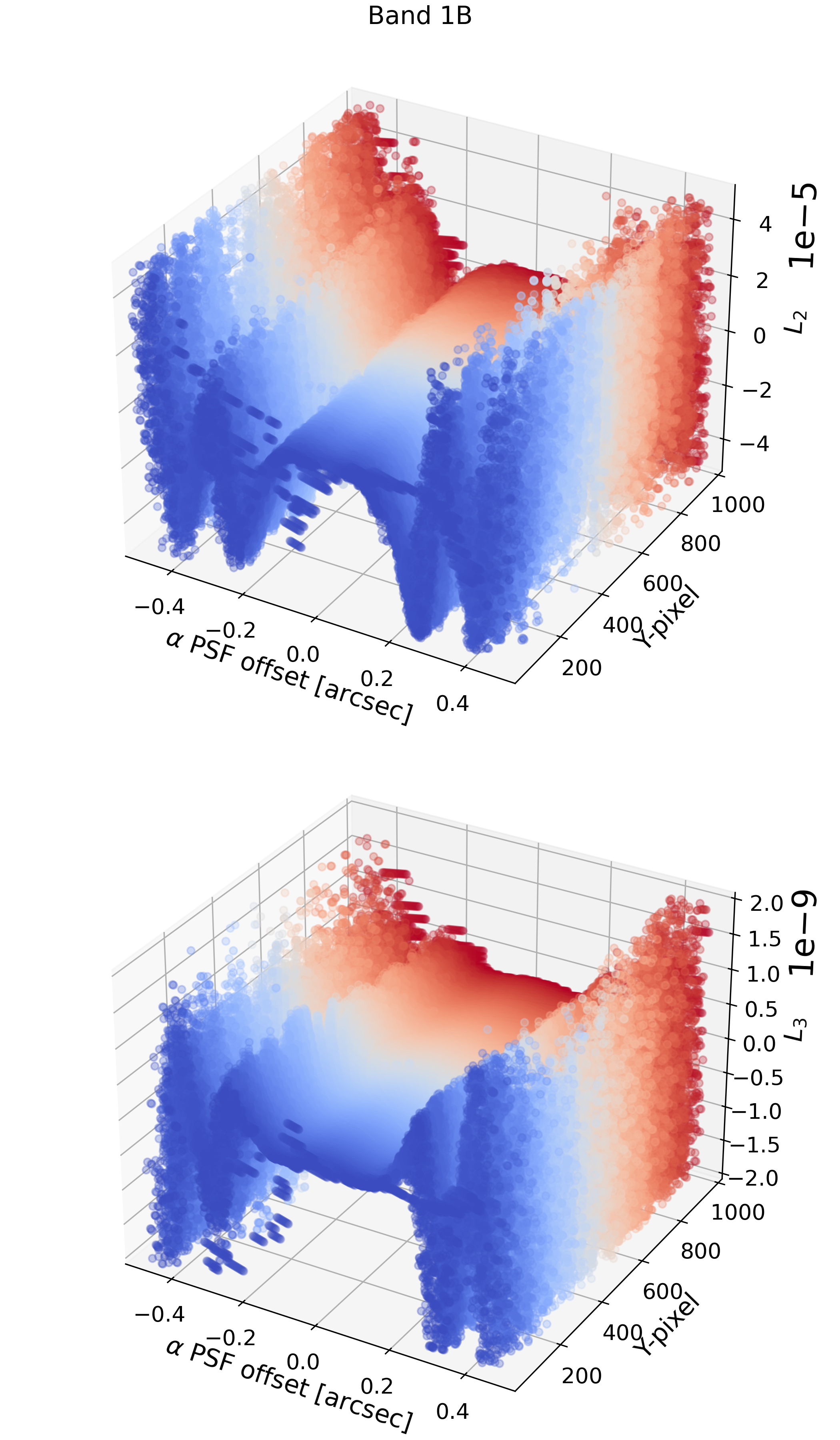} \hfill
 \includegraphics[width=.48\linewidth]{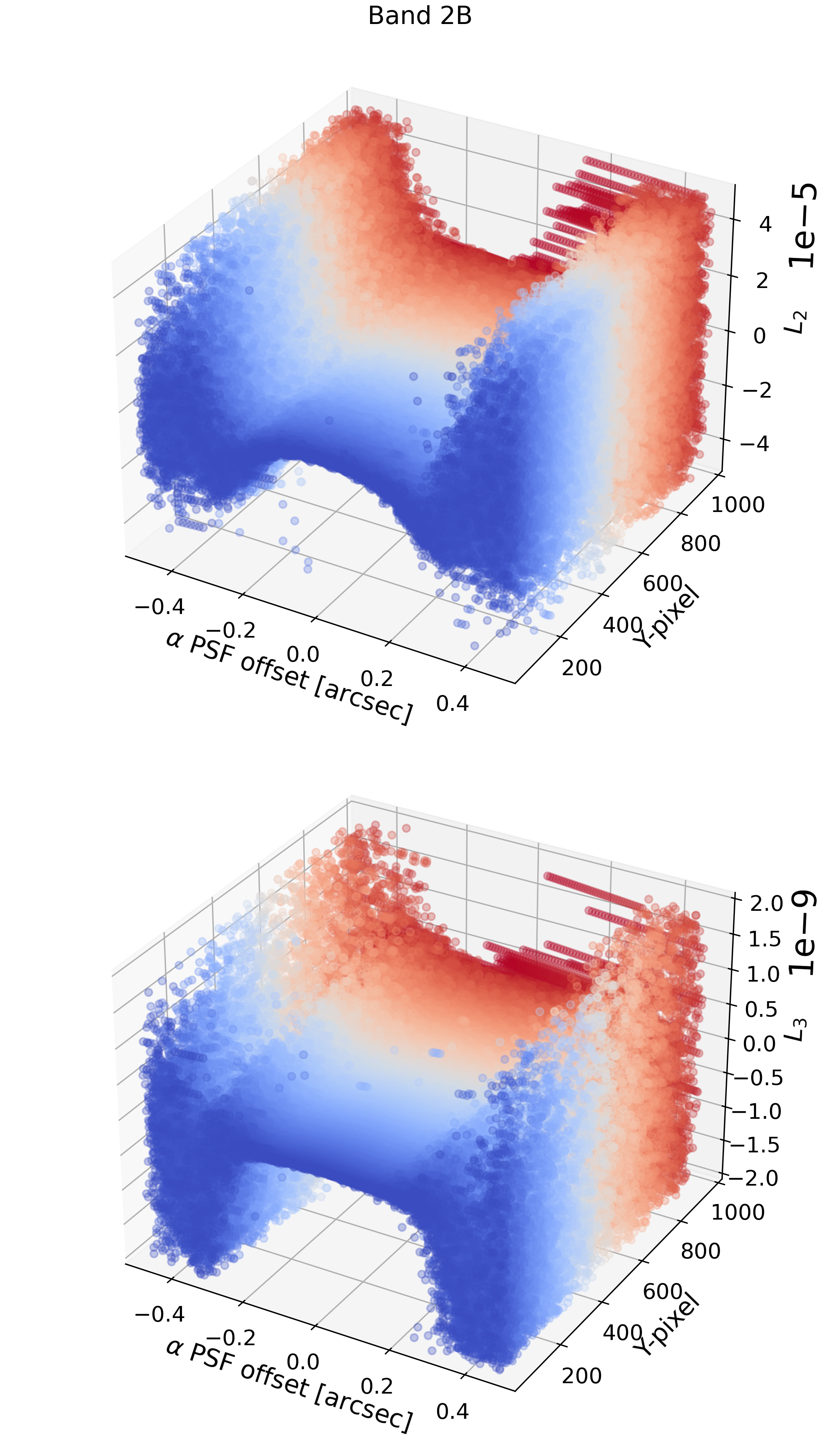}
 \caption{Change in polynomial coefficients $L_2$ and $L_3$ with offset from PSF peak in bands 1B (left) and 2B (right).}
   \label{fig:12b_polyfits}
\end{figure*}

\begin{figure*}
 \includegraphics[width=.48\linewidth]{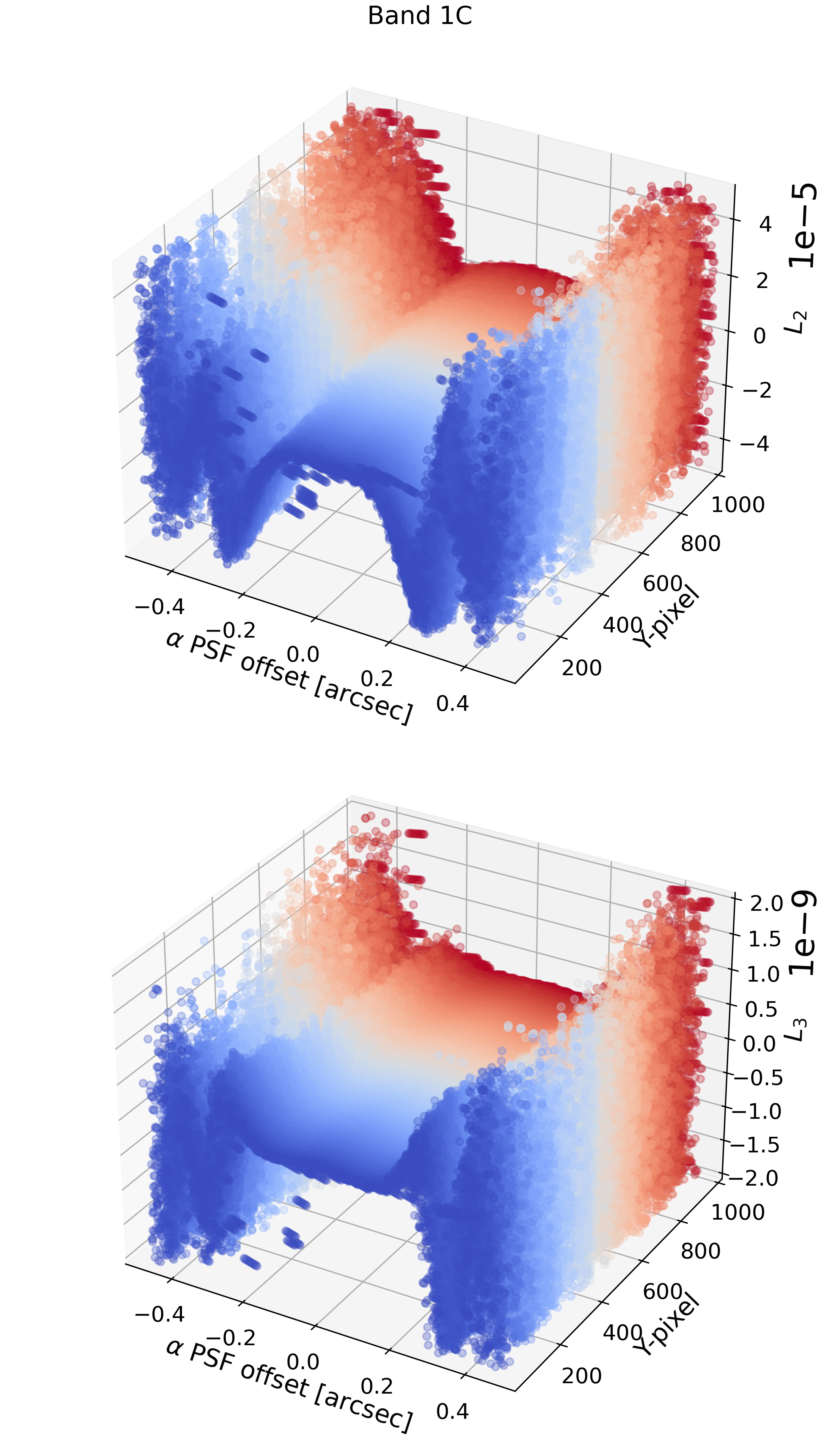} \hfill
 \includegraphics[width=.48\linewidth]{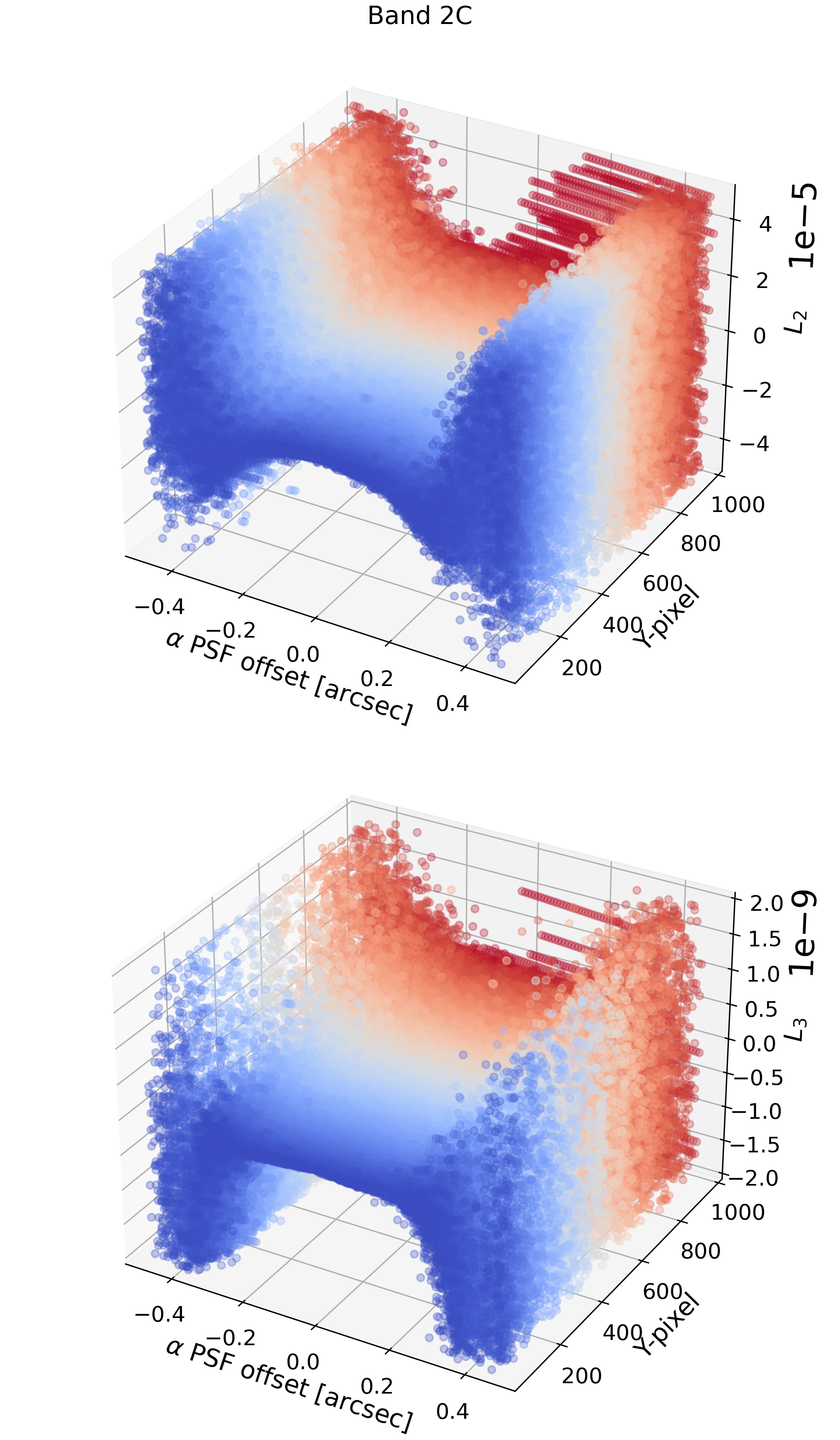}
 \caption{Change in polynomial coefficients $L_2$ and $L_3$ with offset from PSF peak in bands 1C (left) and 2C (right).}
   \label{fig:12c_polyfits}
\end{figure*}

\end{document}